\documentclass{aastex}
\usepackage{emulateapj5}

\usepackage{color}
\definecolor{darkyellow}{rgb}{.7,.7,.5}
\definecolor{darkgreen}{rgb}{0.,0.6,0.}
\definecolor{darkblue}{rgb}{0, 0, 0.8}
\definecolor{lightgray}{rgb}{0.8,0.8,0.8}
\definecolor{middlegray}{rgb}{0.7,0.7,0.7}
\definecolor{darkgray}{rgb}{0.5,0.5,0.5}
\definecolor{darkred}{rgb}{0.8,0.1,0.1}
\definecolor{darkcyan}{rgb}{0.6,0.8,0.8}
\definecolor{background}{rgb}{0.8,0.8,0.6}
\definecolor{darkmagenta}{rgb}{0.8,0.0,0.1}
\definecolor{text}{gray}{0.2}
\definecolor{heading}{rgb}{0.55, 0, 0}

\shorttitle{$\alpha$ and Heavy elements in bulgelike Stars}
\shortauthors{Pompeia et al.}

\begin{document}
\input{epsf}

\title{Detailed Analysis of Nearby Bulgelike Dwarf Stars III.
$\alpha$ and Heavy-element abundances}

\author{Luciana Pomp\'eia and Beatriz Barbuy}
\affil{Universidade de S\~ao Paulo, IAG, C.P. 3386, 01060-970 S\~ao
Paulo, Brazil }
\email{pompeia@astro.iag.usp.br, barbuy@astro.iag.usp.br}

\and
\author{Michel Grenon}
\affil{Observatoire de Gen\`eve, Chemin des Maillettes 51, CH-1290 Sauverny, Switzerland}
\email{Michel.Grenon@obs.unige.ch}


\begin{abstract}

The present sample of nearby bulgelike dwarf stars has kinematics and metallicities
characteristic of a probable inner disk or bulge origin. Ages derived by using isochrones
give 10-11 Gyr for these stars and metallicities are in the range  -0.80$\leq$
[Fe/H]$\leq$ +0.40. We calculate stellar parameters from spectroscopic data, and chemical
abundances of Mg, Si, Ca, Ti, La, Ba, Y, Zr and Eu are derived by using spectrum synthesis.

We found that [$\alpha$-elements/Fe] show different patterns depending on the
element. Si, Ca and Ti-to-iron ratios decline smoothly for increasing metallicities, and follow
essentially the disk pattern.  O and Mg, products of massive supernovae, and also the
$r$-process element Eu, are overabundant relative to disk stars, showing a steeper decline
for metallicities [Fe/H] $>$ -0.3 dex. [$s$-elements/Fe] roughly track the solar values with
no apparent trend with metallicity  for [Fe/H] $<$ 0, showing subsolar values for the 
metal rich stars. Both kinematical and chemical properties of the bulgelike stars indicate a 
distinct identity of this population when compared to disk stars.

\end{abstract}



\section{Introduction}

The inner stellar populations of the Milky Way are a fundamental key for the understanding of
the chemical and dynamical history of the Galaxy, and they are templates for the study of unresolved
populations in external galaxies. These stellar populations have not been well studied because
of the reddening and dust contamination in the line of sight towards the Galactic center. Due to
the distance and reddening,  basically only luminous objects such as AGB stars, M and K giants and
planetary nebulae have been studied in this region, many of them located in windows of low
extinction as the Baade's Window (BW) and the fields of Blanco \& Terndrup (1989).

The innermost part of the Galaxy comprises a nucleus, within a few hundred parsecs to the center
and  the bulge, a structure of about 2.8$\times$2.1 kpc at wavelength of 12$\mu$m according to
Habing et al. (1985), or a region within a Galactocentric radius R$_{\rm GC}$ $<$ 4 kpc if the
characteristics of globular clusters are considered (Barbuy et al. 1999a). The nucleus of the
Galaxy or the Galactic Center, has a huge molecular cloud, the Central Molecular Zone
(e.g. Serabyn \& Morris 1996; Lis et al. 2001), with intense star formation, and possibly
harboring a black hole (Eckart et al. 2001; Ghez et al. 2000). The bulge is a more extended
and sparse component than the Galactic center, with high velocity dispersion, 
$\sigma$ $\sim$ 100 kms$^{-1}$ (e.g. Sadler et al. 1996), and sustained by rotation
(e.g. Wyse \& Gilmore 1992). Studies of the COBE - DIRBE data and other sources showed that
the Milky Way contains a central bar (e.g. Blitz \& Spergel 1991; Weiland et al. 1994; Wada
et al. 1994; Nikolaev \& Weinberg 1997), although it is not clear if bulge and bar are the
same structure or share the same stellar populations (e.g. Kuijken 1996; Combes 1998; Gerhard
2000; van Loon et al. 2003 and references therein). The outer border of the bulge, where
it connects to the inner disk, and the inner disk itself, still have a poorly known structure
(Wyse et al. 1997; Gerhard 2000).

As the center of the gravitation potential well, the inner Galaxy is populated with stars
in transit, born in halo and disk environments. Within
the Galactic Center, M supergiants, OH-IR, luminous AGB and Wolf-Rayet stars were identified
(e.g. Catchpole et al. 1990; Krabbe et al. 1995; Blum et al. 1996; Frogel et al. 1999). Only
recently high-resolution metallicities have been reported for some of these objects (Ramirez et
al. 2000; Carr et al. 2000), indicating a mean metallicity near the solar one. Age measurements
suggest that a fraction of stars of the Galactic center was formed in the last 3-7 Gyr
(Tamblyn \& Rieke 1993; Krabbe et al. 1995; Frogel et al. 1999).

A few studies of the bulge population have been published during the last two decades
(e.g. Whitford \& Rich 1983; Rich 1988;
 McWilliam 1997; Wyse et al. 1997; Wyse 2000; and references therein). Age determinations
from HST data have confirmed that most of the bulge stars are old, with $\sim$ 13 Gyr (Ortolani
et al. 1995; Feltzing \& Gilmore 2000; Ortolani et al. 2001). The metallicity distribution of
this population ranges from [Fe/H] $\simeq$ -1.6 to + 0.55 dex (McWilliam \& Rich 1994,
hereafter MR94) with a mean of
 [Fe/H] $\approx$ -0.25 or [M/H] $\approx$ 0 (MR94; Zoccali et al. 2003). MR94 and Rich \&
McWilliam (2000, hereafter RM00) performed a detailed abundance analysis of a sample of K giants
in the BW. Abundances of $\alpha$, $r$ and $s$-process elements were derived revealing some
peculiarities, with $\alpha$-elements showing different behaviors depending on the element:
O, Mg and Ti are overabundant, while Si and Ca show a trend similar to disk stars. $s$-process
elements track roughly the behavior of disk stars while the $r$-process element Eu is enhanced.

Very few data were published about inner disk stars and the metallicity distribution of this
region has not been investigated in detail. One high-resolution abundance analysis was performed
for a sample of B stars showing enhanced chemical abundances relative to solar neighborhood for
$\alpha$-elements, except for oxygen (Smartt et al. 2001).

In the present work we derive chemical abundances for a sample of 35 stars pertaining to a
population with peculiar kinematics. It consists of stars with highly eccentric orbits,
pericentric distances R$_{\rm p}$ $<$ 5.5 kpc, and absolute vertical velocities of less than
20 kms$^{-1}$. Isochronal ages of 10-11 Gyr were inferred for this population
 (Grenon 1998). The metallicity
distribution is identical to that of the bulge sample studied by MR94 (Grenon 1999). The details
on kinematical properties and theories about the origin of such population are described in
Pomp\'eia, Barbuy \& Grenon (2002a, hereafter Paper I) and in a series of papers by Grenon
(1990, 1999, 2000), and Castro et al. (1997).  The main goal of this work is to analyse a larger
sample of these stars in order to improve our knowledge on the chemical distribution of stars from
the inner Galaxy.  These results could shed some light on the following questions:
(1) what is the abundance pattern of this group?; (2) are the chemical distributions of this
population similar to the population of any other component of the Galaxy?; (3) given the
chemical distributions, what can be inferred about the origin of this population?


\section{Observations}

Sample stars were observed at the 1.52m ESO telescope at La Silla, Chile, in September 1999.
The spectra were obtained using the FEROS spectrograph, with coverage from 356 to 920 nm and
resolution of R  48,000. Most spectra have signal to noise ratios S/N $>$ 100. Reductions were
performed using the DRS (online data reduction system of FEROS) and for a subsequent reduction,
tasks of the IRAF package were applied (see also Paper I).


\section{Analysis}


\subsection{Stellar Parameters}

In Paper I, gravities and metallicities were derived enforcing the ionization equilibrium 
of \ion{Fe}{1} and \ion{Fe}{2} lines, but the reliability of such method has been questioned. In  
the atmosphere of late-type stars, Fe I lines are formed under non-LTE (NLTE) conditions, and two main
points have been observed about the derived stellar parameters: gravities measured through 
ionization equilibrium are usually lower than those derived from accurate astrometric values 
(e.g. Fuhrmann et al. 1997; Nissen et al. 1997; Allende Prieto et al. 1999), and metallicities 
calculated applying NLTE corrections are higher than those derived by the ionization imbalance
(Thev\'enin \& Idiart 1999). In order to increase the accuracy of our abundances,
we have reanalyzed the stellar parameters as follows. 

As in Paper I, the effective temperatures were estimated by fitting the H$\alpha$ line profiles; 
trigonometric gravities derived using Hipparcos paralaxes (see Paper I) are adopted; 
microturbulence velocities were inferred requiring that the $\epsilon$(\ion{Fe}{1}) vs. EW 
(equivalent widths) gives no slope; metallicities were calculated from the 
\ion{Fe}{2} lines and checked with metallicities derived from Geneva photometry. MARCS model 
atmospheres (Gustafsson et al. 1975) were employed.  

The reliability of the determination of temperature of cool dwarf stars by Balmer line profiles 
has been discussed in detail in literature (e.g. Furhmann et al. 1993, 1994; Furhmann 1998; 
Cowley \& Castelli 2002), even for stars as cool as T$_{\rm eff}$ $\sim$ 4800 K (Barklem et al. 2002), 
showing that the involved errors are small. 

In Table 1 the inferred stellar parameters are reported. Metallicities derived from both Fe I 
and Fe II lines are given, together with the standard deviations. Fe I and Fe II line lists and 
EW are given in Paper I. We have improved our Fe II line list with 5 lines, 6149.25, 6179.40,
7222.39, 7224.46 and 7449.33 $\rm \AA$  from Chen et al. (2003). As expected, metallicities derived from 
Fe I give smaller values than those derived from Fe II (e.g. Thev\'enin \& Idiart 1999). HD 11306
show a large difference between metallicity values and a high standard deviation for Fe II lines. 
For this star we chose [FeI/H] = -0.63 which yields the best fit to the stellar spectrum.      
The uncertainties of the fundamental parameters are: $\sigma$(T$_{\rm eff}$) = 50 K; 
$\sigma$(log $g$) = 0.15 dex, $\sigma$([Fe/H]) = 0.10 dex; and $\sigma$ $\xi_{t}$ = 0.10.

\placetable{tb1}


\subsection{Line List}

Atomic lines are selected based on the paper of MR94, and improved with lines of Barbuy et al. (1999b, 2003) 
and Smith et al. (2000). Oscillator strengths, adopted from the cited papers and other
sources (see Table 2), are tested by fitting the synthetic to the observed line profiles in the
Solar Flux Atlas of Kurucz et al. (1984). The solar abundances are from Grevesse et al. (1996).
Synthetic lines with a very good match to the solar spectra are chosen, with the exception of one
barium line, which is discussed below. In Table 2 we give the atomic line list for the $\alpha$ and
heavy-element lines used in this work. 

Oxygen abundances derived in Paper I are recalculated using the new stellar parameters and 
taking into account the Ni I line at 6300.34 $\rm \AA$ with log gf = -2.31 (Allende Prieto 
et al. 2002). For the solar abundance we adopted [O/Fe] = 8.77.  
Hyperfine structures (HFS) for the Eu I  6645.41 $\rm \AA$
and the \ion{Ba}{2} 4554.03, 4934.10 and 6496.41 $\rm \AA$ lines are taken into account following
Fran\c cois (1996 and private communication) and Hill et al. (2000). Below we discuss the selected
line list.

\placetable{tb2}

\subsubsection{$\alpha$-elements}

The four selected Ca I lines are relatively intense and were well fitted to the solar spectrum.
Six Ti I lines were chosen, with three moderately weak lines at 6303.77, 6312.24 and 6336.11
$\rm \AA$. Such lines are near to the Fe I features at 6303.46 and 6336.83 $\rm \AA$, but we found a
good separation between lines thanks to the high resolution and good quality of the spectra.
The Ti I line at 6554.22 $\rm \AA$ is also near the H$\alpha$ profile, which required a
careful definition of the continuum near lines.

Three of the five Si I lines adopted, at 6237.33, 7405.79 and 7415.96 $\rm \AA$, are moderately intense.
Although the two last features are blended to CN and Cr I lines, the spectrum synthesis yielded a
good match to the solar spectrum. The remaining two Si I 6721.84, and 7226.21 $\rm \AA$ lines are
weak but apparently free of blendings. Only the line 6319.24 $\rm \AA$ from the magnesium
triplet at 6316 - 6319 $\rm \AA$ showed a good fit to the solar spectrum. The second Mg I line adopted,
at 7387.70 $\rm \AA$, has features of Co I near the blue wing, and Mg I in the red, but we found a
good agreement between solar and synthetic features.

\subsubsection{Heavy elements}

Both La II lines are very small features even if free of blendings. Therefore most of the derived
abundances for this element represent upper limits. Also the adopted zirconium  Zr II 6114.85
$\rm \AA$, and Zr I 6134.57, 6140.46 and 6762.40 $\rm \AA$ lines are very small. The line at
6140.46 $\rm \AA$ is near the strong Ba II feature at 6141.73 $\rm \AA$, which required a more
careful fit of the continuum. In Figure 1 we show the synthetic and observed spectra of this
region for HD 152391. Yttrium and europium are represented in our sample by the single weak lines
Y I 6795.41 and \ion{Eu}{2} 6645.10 $\rm \AA$ respectively. The Y I line is near the Fe I 6795.80
$\rm \AA$ feature, but no blendings were observed. Y, Zr and La show very small features in
the spectra of the stars (see Tables 4 and 5), and hyperfine structures were not taken into account
in the abundance analysis.  

\begin{figure}
\label{fg1}
\includegraphics[scale=.3, angle=-90]{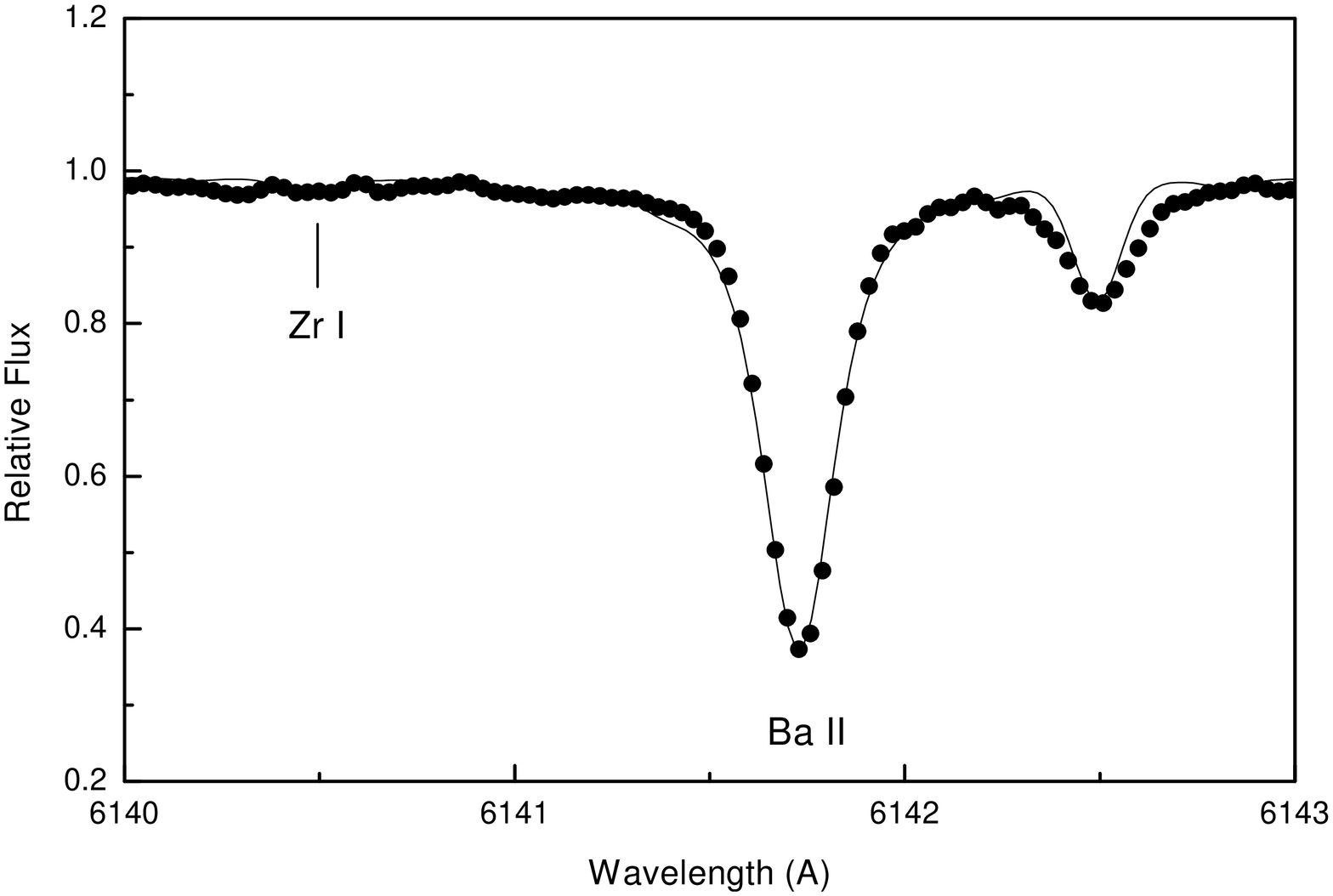}
\caption{Synthetic fit in the region of the Zr I 6140.46 $\rm \AA$ and Ba II 6141.73 $\rm \AA$
lines for the star HD 152391.}
\end{figure}

\placefigure{fg1}

\par {\it Barium, a special case:}
barium lines are very intense and with many isotopic components, some subject to hyperfine splittings.
Although many isotopes are present in the resonant lines, the isotopic shifts are very small,
$\leq$ 2 m$\rm \AA$ (Mashonkina et al. 1999). Three Ba II lines are adopted to infer the barium
abundances: the resonance lines at 4554.03 and 4934.10 $\rm \AA$, and one line in the red,
6496.91 $\rm \AA$. Atomic HFS components of Fran\c cois (1996 and private communication), based on
the HFS of Rutten (1978), are adopted.

In the synthesis of the solar spectrum, where the barium lines are saturated, we found a non-solar
barium abundance, [Ba/Fe] = +0.22, for the 6496.91 $\rm \AA$ line (see sect. 3.3.2). We also found
that the core of the observed profile of this line and the two resonant lines are slightly deeper
than the synthetic ones. Due to the saturation, the core of these lines form in a shallower atmospheric
layer, under NLTE conditions. Problems with abundance analysis of barium lines  have long
been discussed in the literature (e.g. Rutten 1978; Magain 1995; Sneden et al. 1996; McWilliam 1997;
Mashonkina et al. 1999, Mashonkina \& Gehren 2000), particularly involving the fitting of the core of
the solar Ba II lines, as described by Mashonkina et al. (1999) and Mashonkina \& Gehren (2000),
and non-solar abundances have been reported for the synthesis of these lines (e.g. Smith et al. 2000).


\subsection{Abundances}

LTE stellar abundances of Mg, Si, Ca, Ti, Y, Zr, Ba, La and Eu are derived by fitting the
synthetic to the observed spectra. The spectrum synthesis code is described in Cayrel et
al. (1991) and Barbuy et al. (2003). The synthetic spectra were convolved with  
Gaussian profiles with FWHM from 0.10 to 0.15 $\rm \AA$. Molecules of MgH 
(A$^{2}\Sigma-\rm X^{2}\Pi$),
C$_{2} (\rm A^{3}\Pi-\rm X^{3}\Pi)$, CN blue ($\rm B^{2}\Sigma-\rm X^{2}\Sigma$),
CH ($\rm A^{2}\Delta-\rm X^{2}\Pi$), CH ($\rm B^{2}\Delta-\rm X^{2}\Pi$), CN red
($\rm A^{2}\Pi-\rm X^{2}\
Sigma$) and TiO $\gamma$ ($\rm A^{3}\Phi-X^{3}\Delta$) are included. In Figure 2
we show the fit to the Ca I 6455.60 $\rm \AA$ line (top) and the Ba II 6496.91 $\rm \AA$
line (bottom) for HD 10758. In Table 3 we report the derived abundances along with the scatter
(standard deviation) among different lines. In Tables 4 and 5, equivalent widths are given for
each line.

\placetable{tb3}

\placetable{tb4}

\begin{figure}
\label{fg2}
\rotatebox{270}{\resizebox{!}{8cm}
{\includegraphics{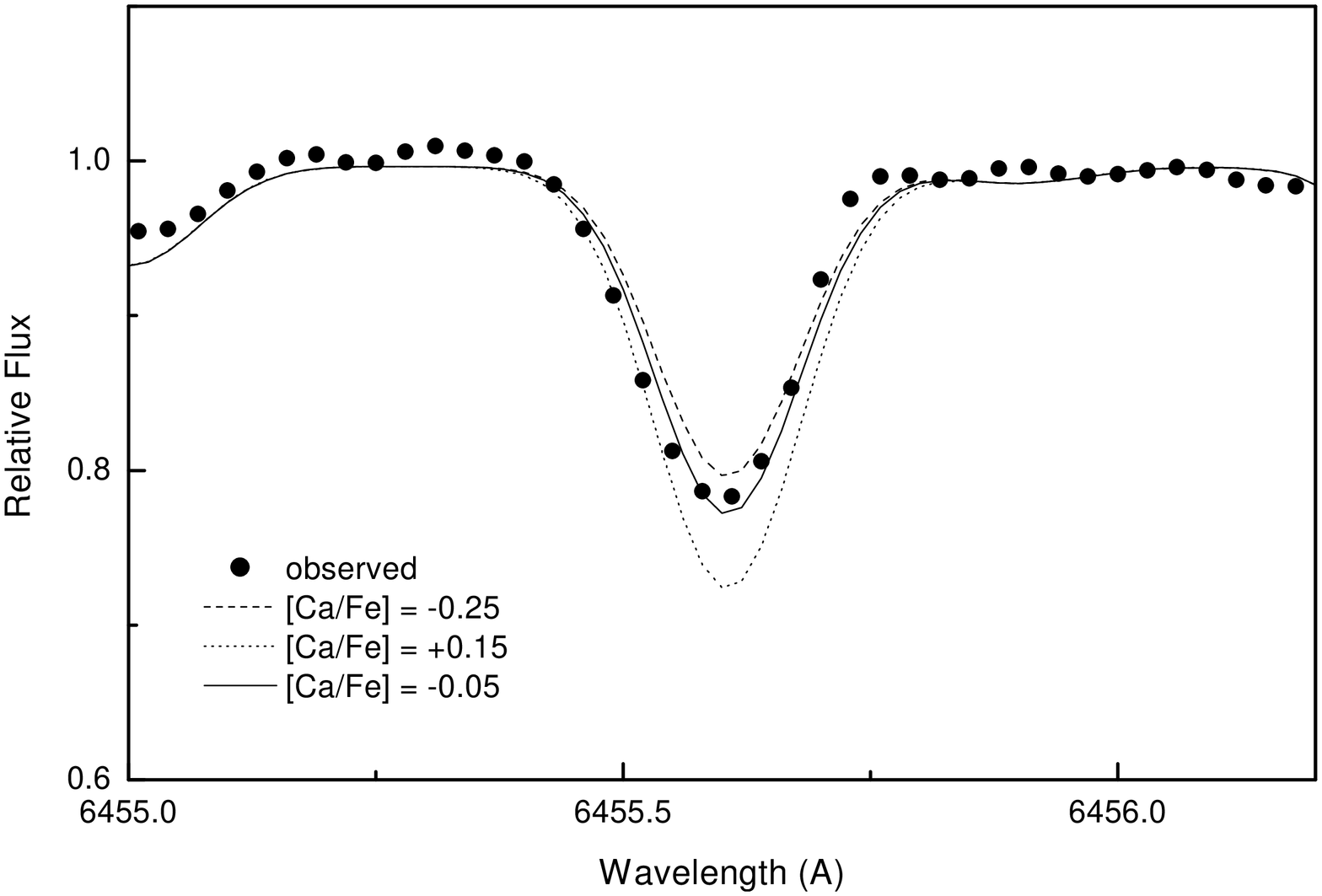}  \includegraphics{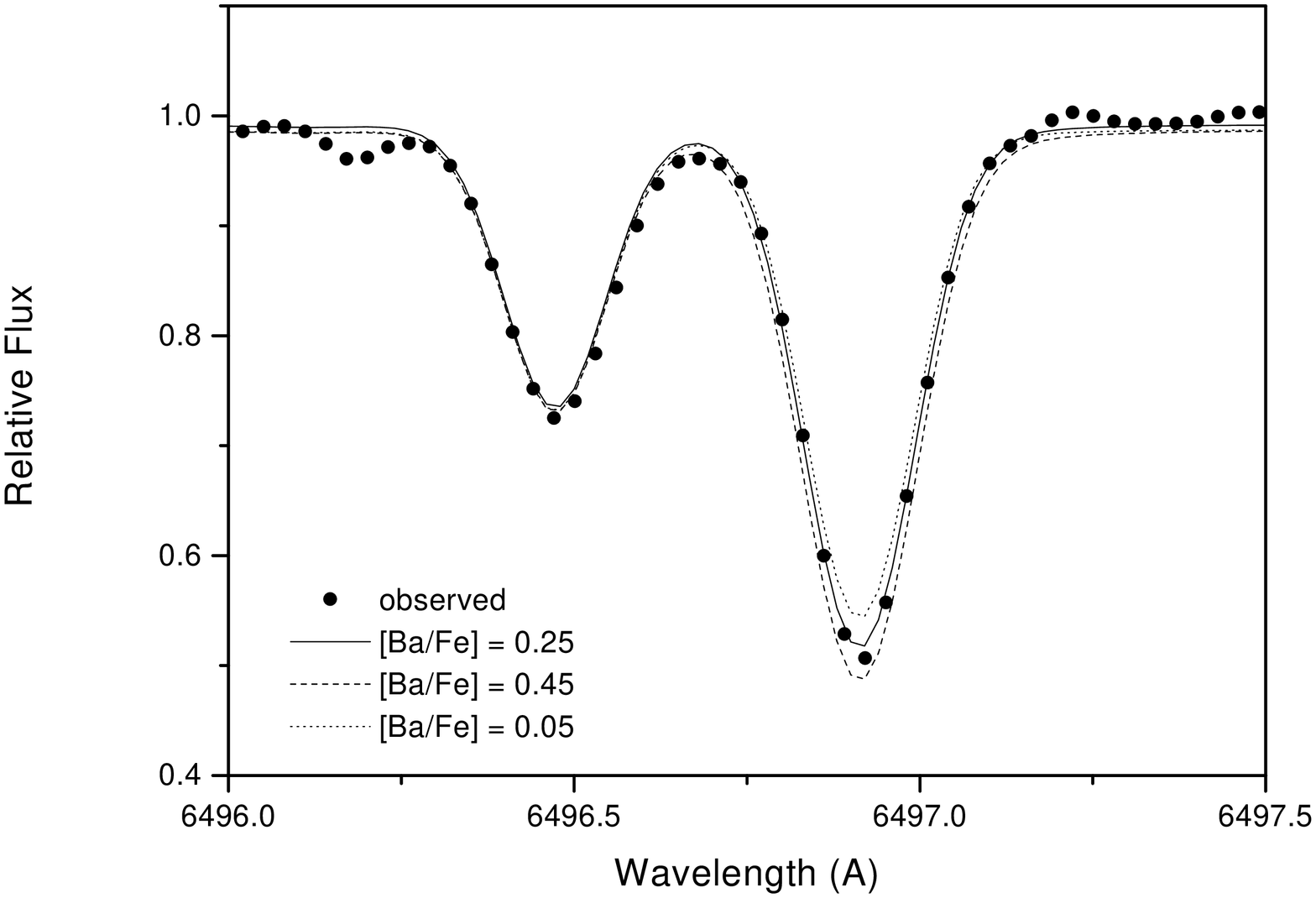}}}
\caption{Examples of the fitting procedure for HD 10758. Observed (dots) and synthetic (lines) spectra
for three abundance values (see the legends) around Ca I 6455.60 $\rm \AA$
and Ba II 6496.91 $\rm \AA$ lines. The best fit give [Ca/Fe] = 0.05 and [Ba/Fe] = 0.25. }
\end{figure}

\placefigure{fg2}


\subsubsection{Continuum absorption for $\alpha$-enhanced atmospheres}

The continuum absorption has been computed by taking into account the $\alpha$-element
(O, Mg, Si, Ca and Ti) enhancement for the stars showing [$\alpha$/Fe] $\geq$ 0.2, as being equal to 0.2.
This is important because the $\alpha$-elements are electron donors (magnesium being the most important
element in this respect). With  $\alpha$ enhancement the electron pressure increases, therefore the continuum 
absorption by H$^{-}$ increases. As a consequence, a certain optical depth will be reached at a shallower layer, 
where the gas pressure is lower, therefore the line wings will be less strong.

The continuum is affected in two ways: {\it (i)} molecular bands or a number of atomic lines
 involving the $\alpha$-elements will affect indices in the pseudo-continuum as well as the features; 
{\it (ii)} the true continuum is dominated by H$^{-}$, and therefore it will depend on electron donors, 
among which the $\alpha$-elements Mg and Si are important, whereas Ti is not important in this respect.

\subsubsection{Abundance Errors}

Errors in derived abundances depend on many factors. The uncertainties in stellar parameters are an
important error source. To evaluate the effect of these uncertainties, we changed the stellar parameters
of HD 143016 and HD 10576 by $\Delta$ T$_{\rm eff}$  +50 K, $\Delta$ log $g$ = +0.15 dex,
$\Delta\xi$ = +0.10 and [Fe/H] = 0.10 dex, and calculated the abundance variations. 
Abundance changes are summarized in Table 6, where the total errors, $\delta_{t}$, are also given. 
Table 6 also shows that most abundances vary by less than 0.1 dex, with a maximum value of 0.15 dex.

\placetable{tb5}

The scatter among different lines given in Table 3 are probably due to a sum of factors: uncertainties
in the adopted oscillator strengths and stellar parameters, and due to NLTE and blending effects.
Abundances estimated by using very weak lines as Zr I lines, are subject to errors due to the adoption
of upper limits.

Some of the lines studied here form under departure from LTE, therefore, errors due to non-LTE (NLTE)
conditions should be considered. Zhao et al. (1998) studied NLTE effects on the abundances derived
from Mg I lines. They found a difference of 0.05 dex between NLTE and LTE analysis for the Mg I
6319.49 $\rm \AA$ line. NLTE effects for Mg I lines were also derived for a large sample of dwarf
stars by Shimanskaya \& Mashonkina (2001), where it was found that standard errors for most of the
stars are $\leq$ 0.07 dex. Ca I lines are subject to NLTE effects, however Drake (1991) found that
the errors are negligible for dwarf stars. Thor\'en \& Feltzing (2000) and Thor\'en (2000) found
small changes between LTE and NLTE abundances for Ca I lines in cool dwarf stars, with
T$_{\rm eff}$ $\leq$ 5800 K.

NLTE effects as high as $\sim$0.2 dex were inferred by Mashonkina \& Gehren (2000) for the Ba II
6496.91 $\rm \AA$ line in some cool dwarf stars. In spite of such a high discrepancy, we found a
satisfactory agreement among lines for our sample stars, and all Ba II lines are taken into
account in the abundance analysis. Another error source for the barium lines is the $r$/$s$ solar
mixture choice (Sneden et al. 1996). It affects the isotopic splittings since odd-isotopes are mainly
produced by the $r$-process, while even isotopes, by the $s$-process. But changes of $\sim$ 0.1 dex
or less are expected in abundances due to different isotopic distributions (Burris et al. 2000).

\placetable{tb9}

\section{Results}

In Figure 3 we show the abundance trends for the sample bulgelike stars. Upper limits are 
depicted by down triangles.  It seems from these plots that the $\alpha$-elements have different 
behaviors depending on the element, and can be roughly divided into two groups. The first group 
includes Si, Ca and Ti, and the second, O and Mg, which have similar pattern (except for the upper 
limit values) to the $r$-process element Eu. In the following sections we describe the derived distributions 
and the formation theories of these elements.

\subsection{ Si, Ca and Ti}

Si and Ca are prototypical $\alpha$-elements, synthesized during hydrostatic burning of massive stars
(Woosley \& Weaver 1995), with a small contribution by Type Ia supernovae (SN Ia) (Tsujimoto et al. 1995).
The typical enrichment timescale of the interestellar medium (ISM) due to massive stars
is of some $\sim$ 10$^{6}$ yr. The formation efficiency of both  Si and Ca depends on the mass of the
progenitor of the Type II supernova (SNe II). The production factor of these elements are larger for
a $\approx$ 25 M$_{\odot}$ progenitor, while declining factors are predicted for lower or higher
mass stars (see Figure 6 of RM00). Nucleosynthetic prescriptions associate Ti formation to the iron peak
elements (Thielemann et al. 1996), but the observations strongly suggest connections with the
$\alpha$-elements (e.g. Wheeler et al. 1989; Edvardsson et al. 1993, hereafter Edv93; Chen et al.
2000, hereafter Chen00).

In Figure 3, the [$\alpha$-elements/Fe] vs. [Fe/H] distributions for Si, Ca and Ti are depicted. Such
elements show a monotonic decrease with increasing metallicity for metallicities [Fe/H] $<$ -0.2.
In the metallicity range  -0.2 $<$ [Fe/H] $<$ +0.2, a change in the declining trend, with a flatter
pattern and many subsolar values, is seen. An interesting result from our analysis is the notorious 
similarity between Ca and Ti abundances, with a correlation coefficient of $r$ = 0.89, possibly 
indicating a common nucleosynthetic process.

\subsection{O, Mg and Eu}

O and Mg are produced in massive SNe II ($\sim$ 35 M$_{\odot}$), and these elements are not predicted
in the yields from intermediate to low mass stars (Woosley \& Weaver 1995; Thielemann et al. 1996).
Eu is produced by the $r$-process through rapid neutron captures onto seed nuclei (e.g. Thielemann et
al. 1990). The astrophysical site for this process is still debated (Wallerstein et al. 1995; Hoffman 
et al. 1997; Thielemann et al. 2001) and some possible sites have been suggested: SN II explosion
(e.g. Woosley \& Hoffman 1992; Takahashi et al. 1994), neutron stars merging (Davies et al. 1994;
Rosswog et al. 1999, 2000) and the coalescence of a neutron star and a black hole (Thielemann et al. 2001).

O, Mg and Eu abundances relative to iron in our sample show a slower decrease with increasing metallicity for
-0.8 $\leq$  [Fe/H] $\leq$ -0.2, when compared to Ca, Ti and Si. Beyond this value, a decline is seen followed
by an almost constant trend in the range -0.2 $<$ [Fe/H] $<$ + 0.2, with Mg showing few subsolar values. 
For metallicities above the solar value, Mg and Eu seem to increase with increasing [Fe/H]. [Eu/Fe] show a high 
scatter for all the metallicity range of the sample.

\subsection{$s$-process elements}

[$s$-element/Fe] vs. [Fe/H] distributions are also plotted in Figure 3. The $s$-process nucleosynthesis
consists of slow neutron captures, which occur in intermediate to low mass AGB stars (e.g. K\"appeler et
al. 1989). The efficiency of such process basically depends on: (i) the abundance of the iron-peak elements,
which are the seed nuclei for the $s$-process; (ii) the neutron density at the $s$-process site; (iii)
the $^{13}$C available below the convective envelope at the so called  "$^{13}$C pocket"
- note that the $^{13}$C($\alpha$,n)$^{16}$O is the main source of neutrons in low mass AGB stars
(see Gay \& Lambert 2000); and (iv) the abundance of the neutron poisons, i.e., elements with high cross
sections for neutron capture (Gallino et al. 1998; Busso et al. 2001). As these factors behave
diversely with metallicity, a complex run of the [$s$-element/Fe] ratios with metallicity is
expected (Busso et al. 1999, 2001).

In our sample, the [$s$-element/Fe] values roughly track the solar one, with apparently no trend with
metallicity. Ba shows overabundant values for almost all stars. Y, in opposite way, show many subsolar
values. La and Zr nearly follow the solar abundance with a possible trend of decrease for higher metallicities.
Two stars seem detached from the others in these plots: HD 11397 and HD 14282. HD 11397 has strong $s$-process
lines, and was identified as a CH star by \v{S}leivyte \& Bartkevi\v{c}ius (1990). It has also a high Li
abundance compared to other bulgelike stars (Pomp\'eia et al. 2002b, Paper II) as expected for stars belonging
to binary systems. HD 14282 is moderately enriched in $s$-process elements and has lower Li abundance.

\begin{figure}
\epsscale{1}
\label{fg3}
\plottwo{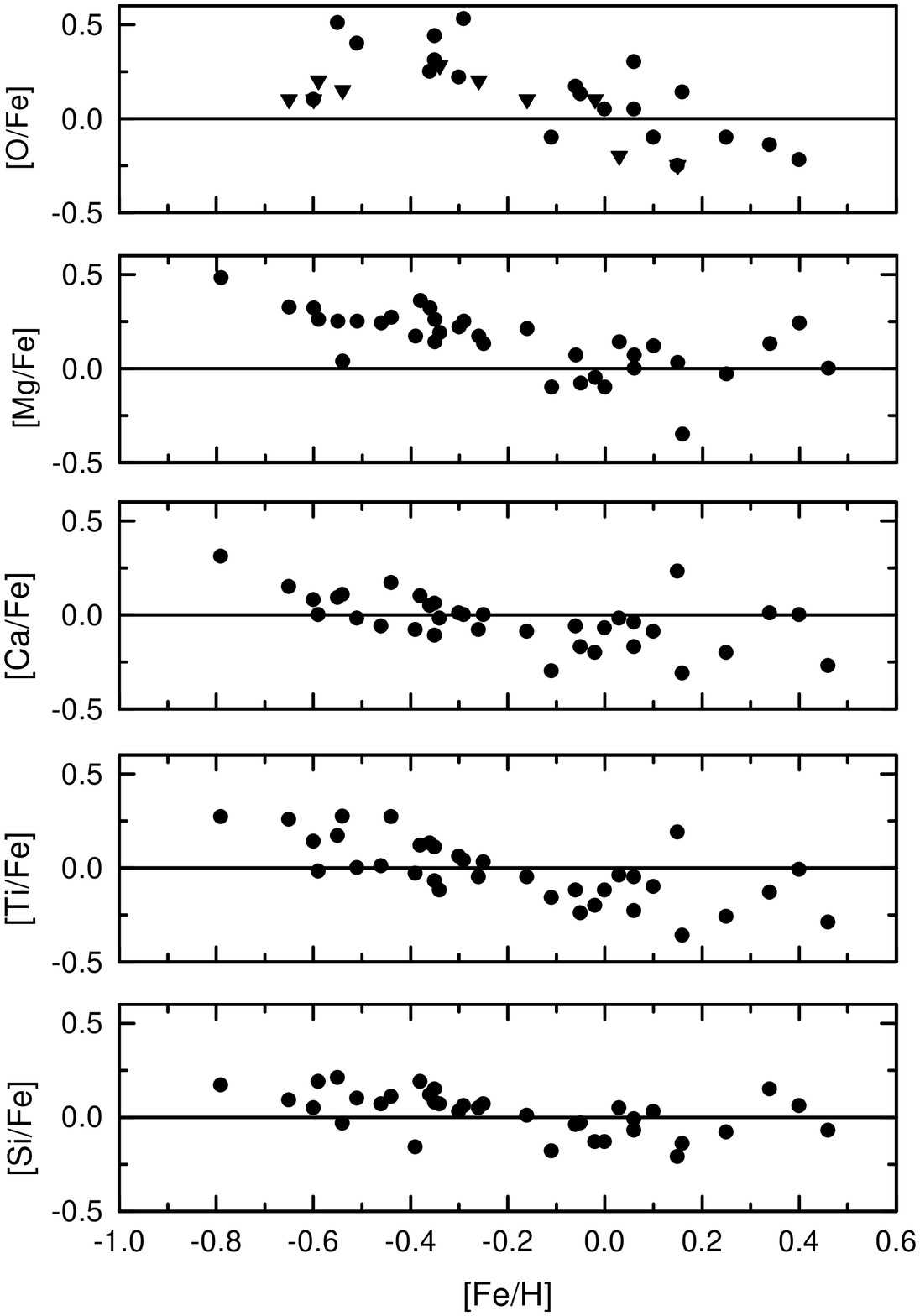}{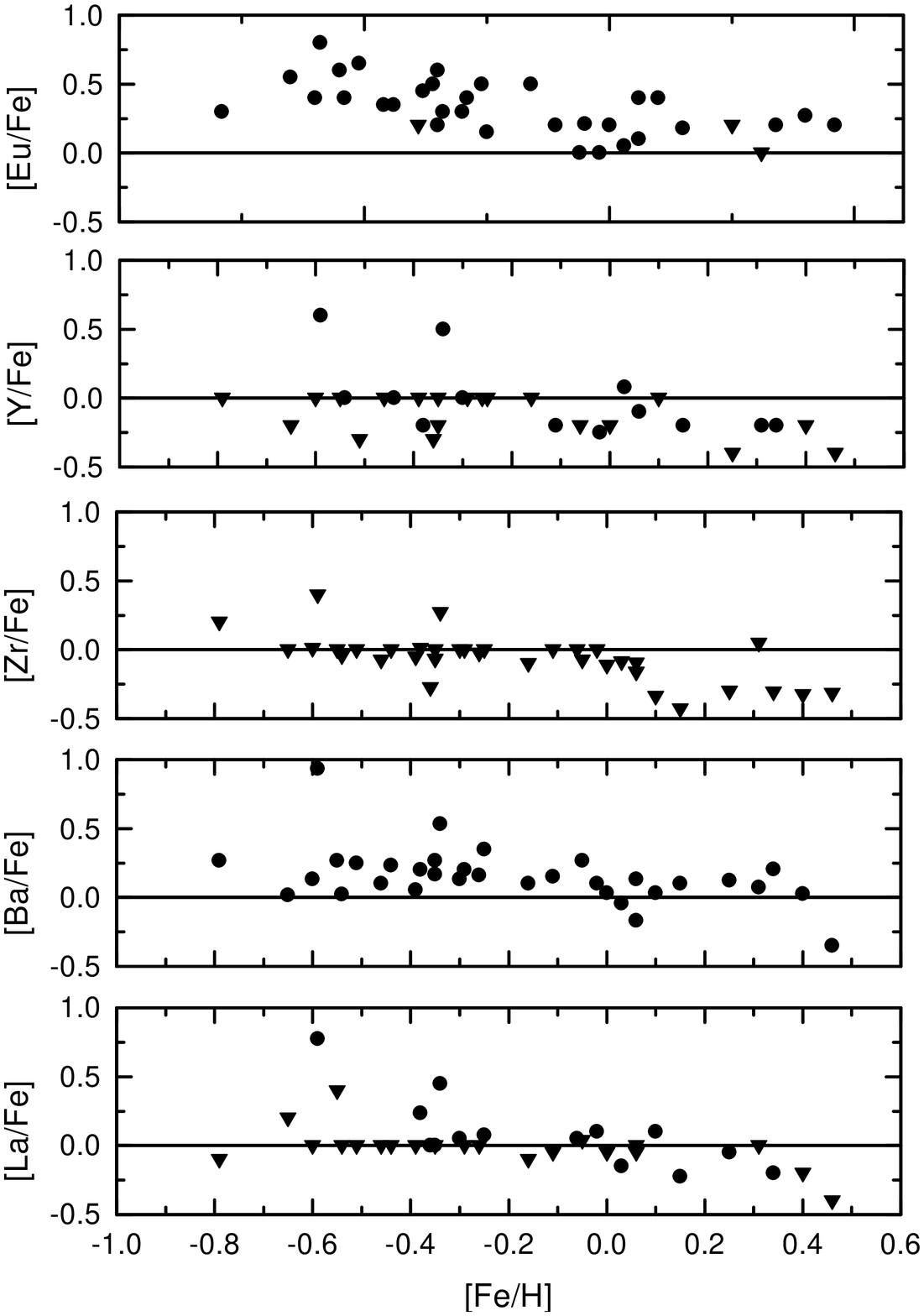}
\caption{[$\alpha$/Fe] (left), [$r$-process/Fe] and [$s$-process/Fe] (right) vs. [Fe/H] relations 
for the bulgelike stars. Downtriangles represent upper limits.}
\end{figure}

\placefigure{fg3}

\section{Comparison to other samples}

In this section we compare the chemical distributions of our sample to previous work on the
same kinematical sample (Barbuy \& Grenon 1990, hereafter BG90; Castro et al. 1997, hereafter Cas97);
to disk samples of Edv93, Nissen \& Edvardsson (1992, hereafter NE92), Chen00, Carretta et al.
(2000, hereafter CGS00),  Feltzing \& Gustafsson (1998, hereafter FG98),
Thor\'en \& Feltzing (2000, hereafter TF00), Smith et al. (2001, hereafter SCK01)
and Koch \& Edvardsson (2002, hereafter KE02); as well as to bulge samples of MR94 and RM00, and results for
the bulge globular cluster NGC 6553 by both Barbuy et al. (1999b) and Cohen et al. (1999).
As discussed before (e.g. Prochaska et al. 2000), the temperature scale of Edv93 has
an offset of $\Delta$T$_{\rm eff}$ $\sim$ +100 K when compared to temperatures derived by
using the Infrared Flux Method (IRFM) and from H$\alpha$ profiles (Gratton et al. 1996). As our
temperatures are inferred from H$\alpha$ profiles, and they agree with IRFM temperatures (Paper I),
we modified the Edv93 data by applying the corrections given in their
Table 6, for a temperature change of $\Delta$T$_{\rm eff}$ = -100 K.


\subsection{Comparison with BG90 and Cas97}

Metal-rich stars ([Fe/H] $\geq$ + 0.2 dex) from the same kinematical sample of our stars have been
studied by BG90 and Cas97. BG90 derived oxygen abundances from the [O I] line at 6300 $\rm \AA$ for a
sample of 11 stars. Cas97 also derived oxygen abundances from this line, and abundances of Mg, Si,
Ca, Ti and Eu among others, for a sample of 9 stars. In Figure 4 we compare our data to BG90 and Cas97.
BG90 found an almost continuous value of [O/Fe] $\approx$  + 0.2 for their sample stars and
Cas97 found a steep decline in oxygen to iron ratios for increasing metallicities. Our results are 
similar to Cas97, although showing a higher scatter. For Mg, Si and Ca, Cas97 values appear as a natural
sequence of our decreasing trends to higher metallicities. For Ti, a steeper decline with 
metallicity is observed in Cas97 compared to our abundance ratios. Our Ti determinations are based on  
six lines, a higher number of lines than that adopted by Cas97, and the trend from our data 
should be reliable. The overabundant behavior of our [Eu/Fe] values appear to be present also for the 
higher metallicities of the Cas97 sample.

\begin{figure}
\epsscale{0.9}
\label{fg4}
\plotone{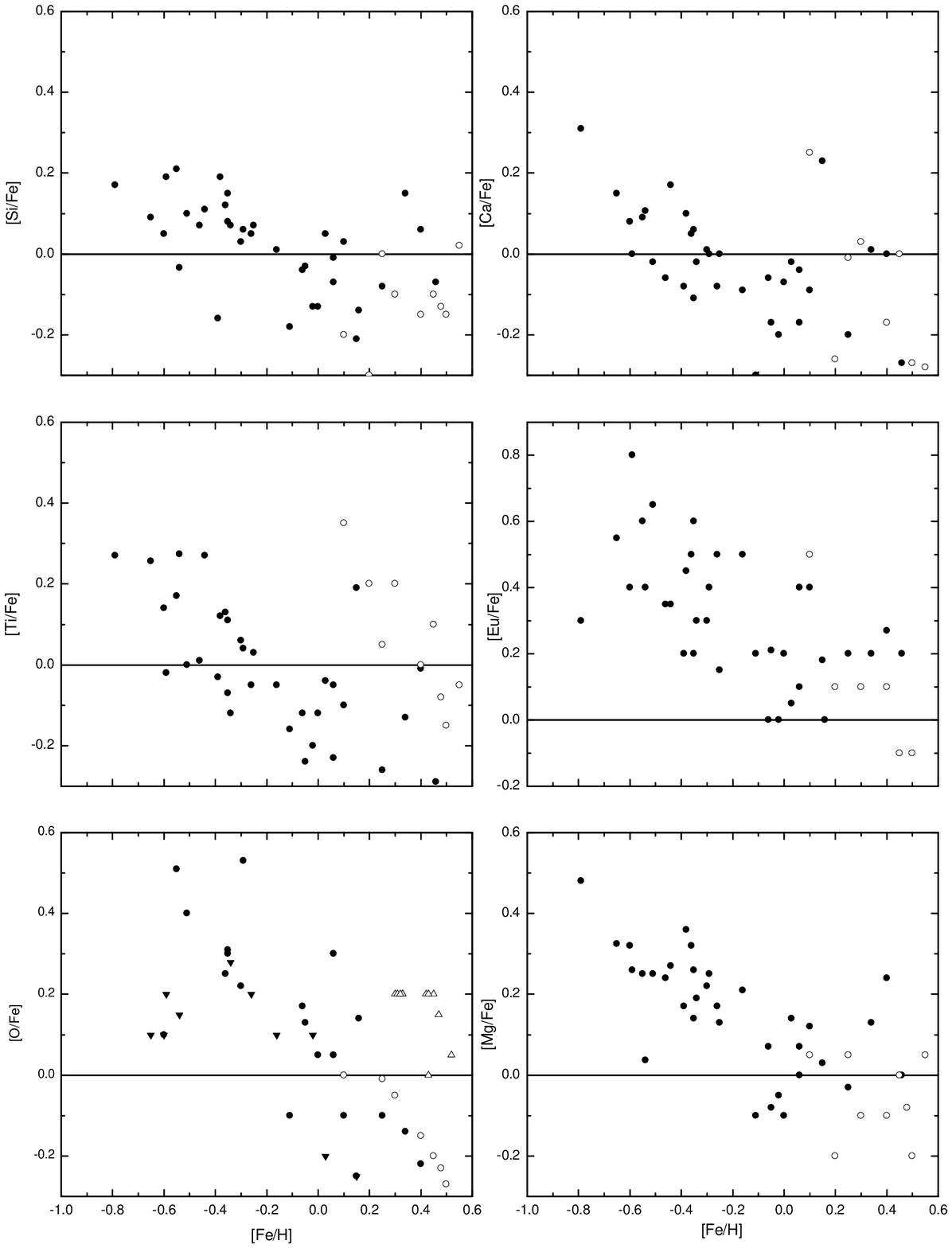}
\caption{[X/Fe] vs. [Fe/H] relations: $\bullet$ present sample compared
with $\circ$ Cas97 and $\triangle$ BG90 samples.}
\end{figure}

\placefigure{fg4}

\subsection{Si, Ca and Ti}

In Figure 5, our sample is compared to disk and bulge results. This figure shows that Si,
Ca and Ti follow the disk trend, with a possible steeper decline for increasing metallicities.
In Figure 6 we compare the mean of  Si and Ca ratios, [$\alpha$'/Fe], to the same ratio of Edv93, for stars with
mean galactocentric radius R$_{m}$ $<$ 7 kpc. We found that the two samples essentially overlap, showing
that the abundances of these elements are similar to those of the disk with smaller R$_{m}$.

The  comparison with the bulge stars shows that our sample show similar values of [Ca/Fe]. On the other hand, 
the bulge shows higher [Si/Fe] and even higher [Ti/Fe] values, with no apparent declining trend with 
metallicity. The results from both Barbuy et al. (1999b) and Cohen et al. (1999) for NGC 6553
show higher Si, Ca and Ti abundances, even if the [Fe/H] value found for the cluster differ in
these studies ([Fe/H] = -0.55 from Barbuy et al. and [Fe/H] = -0.17 from Cohen et al.).

For metallicities above solar, the samples of Edv93, FG98 and TF00 suggest a roughly constant
[Si/Fe] ratio and [Ca/Fe] with a possible decreasing trend. But such trends are not confirmed by 
Chen00 and Cas97. Our sample, as well as bulge stars from MR94, have a larger scatter with no
well-defined trend. For Ti, results from Edv93 show a slightly decreasing pattern, and the opposite
behavior is seen in Chen00's sample.

\begin{figure}
\small
\rotate
\epsscale{0.7}
\label{fg5}
\plotone{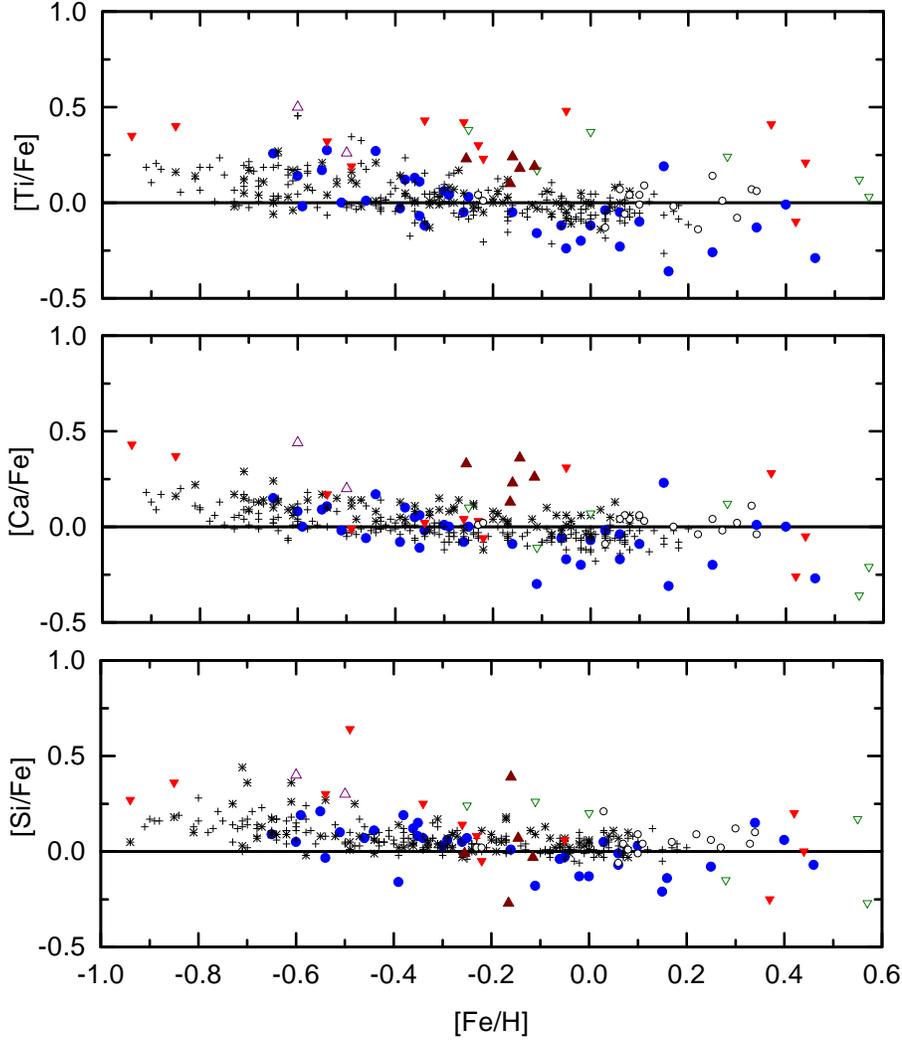}
\caption{[$\alpha$/Fe] vs. [Fe/H] relations for Ti, Ca and Si: present sample (filled circles),
Edv93 (crosses), Chen00 (stars), TF00 (open circles), MR94 (filled downtriangles),
and RM00 (open downtriangles), Cohen et al. (1999) (filled triangles), and Barbuy et al.
(1999) (open trianlges).}
\end{figure}

\placefigure{fg5}

\begin{figure}
\label{fg6}
\includegraphics[scale=.4, angle=0]{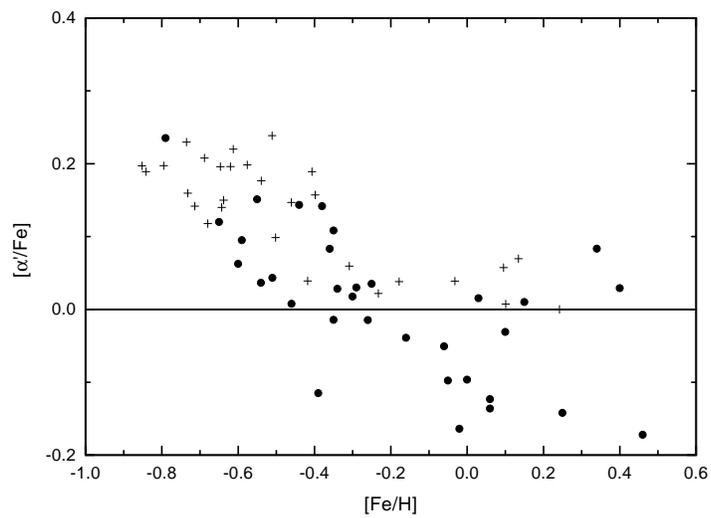}
\caption{[$\alpha$'/Fe] vs. [Fe/H] relation: $\bullet$ present sample, + Edv93.}
\end{figure}

\placefigure{fg6}

\subsection{O, Mg and Eu}

Compared to results for disk stars of NE92, CGS00, SCK01 and KE02, the present sample shows an enhanced 
pattern for Eu for almost all the metallicity distribution (Figure 7), while Mg and O abundance ratios
(when oxygen upper limits are not taken into account) are among the highest values of the disk
for [Fe/H] $<$ -0.1. Relative to bulge stars, an overlap is apparent for [Eu/Fe] and [O/Fe], with 
lower [Mg/Fe] values for the bulgelike stars.

There is a contradictory behavior between O and Mg for supersolar metallicities in NE92, Edv93,
FG98 and CGS00 samples. While [Mg/Fe] shows an upturn with increasing trend, [O/Fe] seems to decrease 
with increasing metallicities. According to GGS00, such behavior suggests that the O/Mg ratio produced 
in SN II vary as a function of metallicity. The bulge stars seem to follow those trends.
Eu observations of disk stars of higher metallicities  show  conflicting results:
while Tomkin et al. (1997) and FG98 found that [Eu/Fe] values increase with metallicity,
Woolf et al. (1995), Cas97 and Mashonkina \& Gehren (2000) samples point to a decreasing trend.
Our data seem to agree with the first group. In the high metallicity
range we have very few data and the behavior analysis in this range is not possible.

\begin{figure}
\epsscale{0.7}
\label{fg7}
\plotone{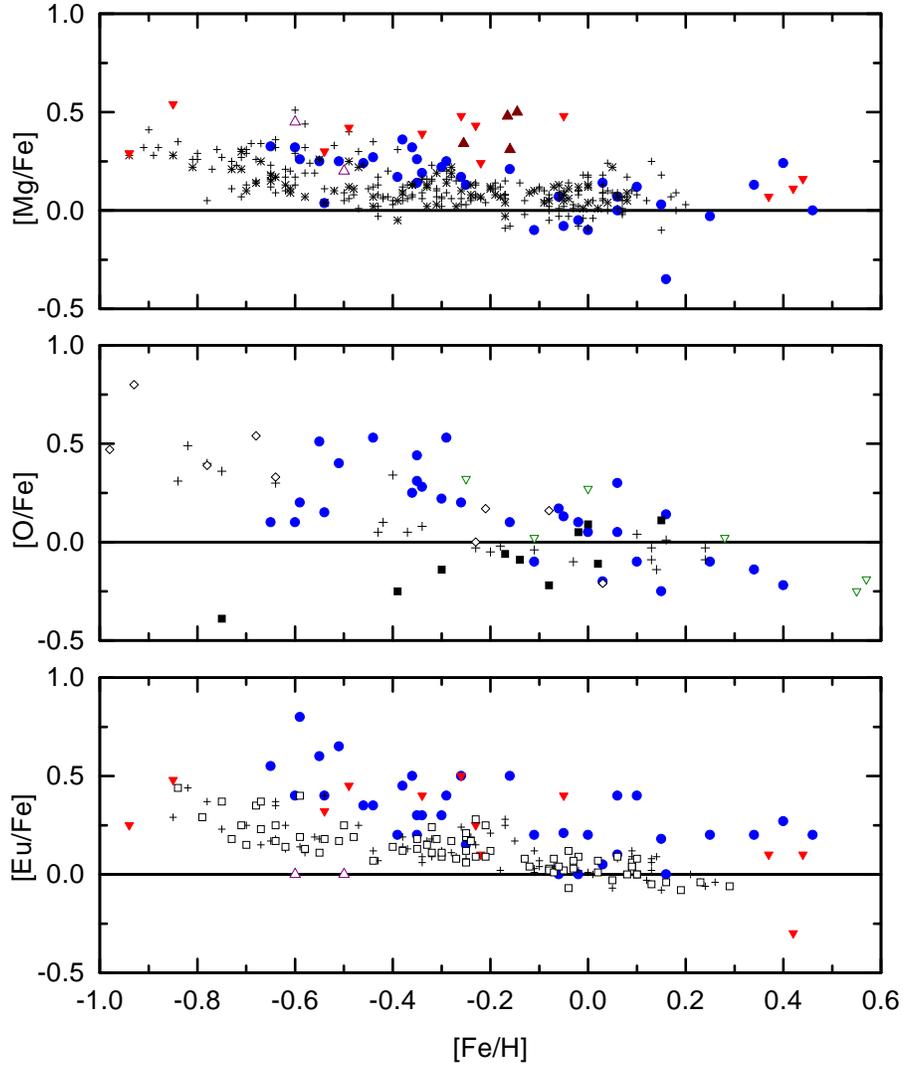}
\caption{[$\alpha$/Fe] vs. [Fe/H] relations for Mg, O and Eu. Symbols
as in Figure 5 except for oxygen, for which we adopted Nissen \&  Edvardsson (1992) (crosses), and more data from:
Carretta et al. (2000) (open diamonds), Smith et al. (2001) (filled squares), and Koch \& Edvardsson (2002) (open
squares) (see text).}
\end{figure}

\placefigure{fg7}

\subsection{$s$-process elements}

Fewer data are available in the literature for the s-process elements. In particular,
La has very few data in the present metallicity range, and in Figure 8 this element is compared
only to the bulge sample of MR94.
As can be seen from this Figure, Zr follows the disk distribution with a lower mean value. Compared to
bulge stars of MR94, our [Zr/Fe] values are enhanced, although MR94 decided not to use their
Zr abundances. Y and Ba distributions
for the present sample overlap those of disk stars, with [Ba/Fe] values comparable with  the highest
values of the disk and apparently following
the bulge pattern, while our [Y/Fe] values are comparable with the lowest disk and bulge values. For La a good
match between our distribution and the bulge sample is found.

\begin{figure}
\epsscale{0.7}
\label{fg8}
\plotone{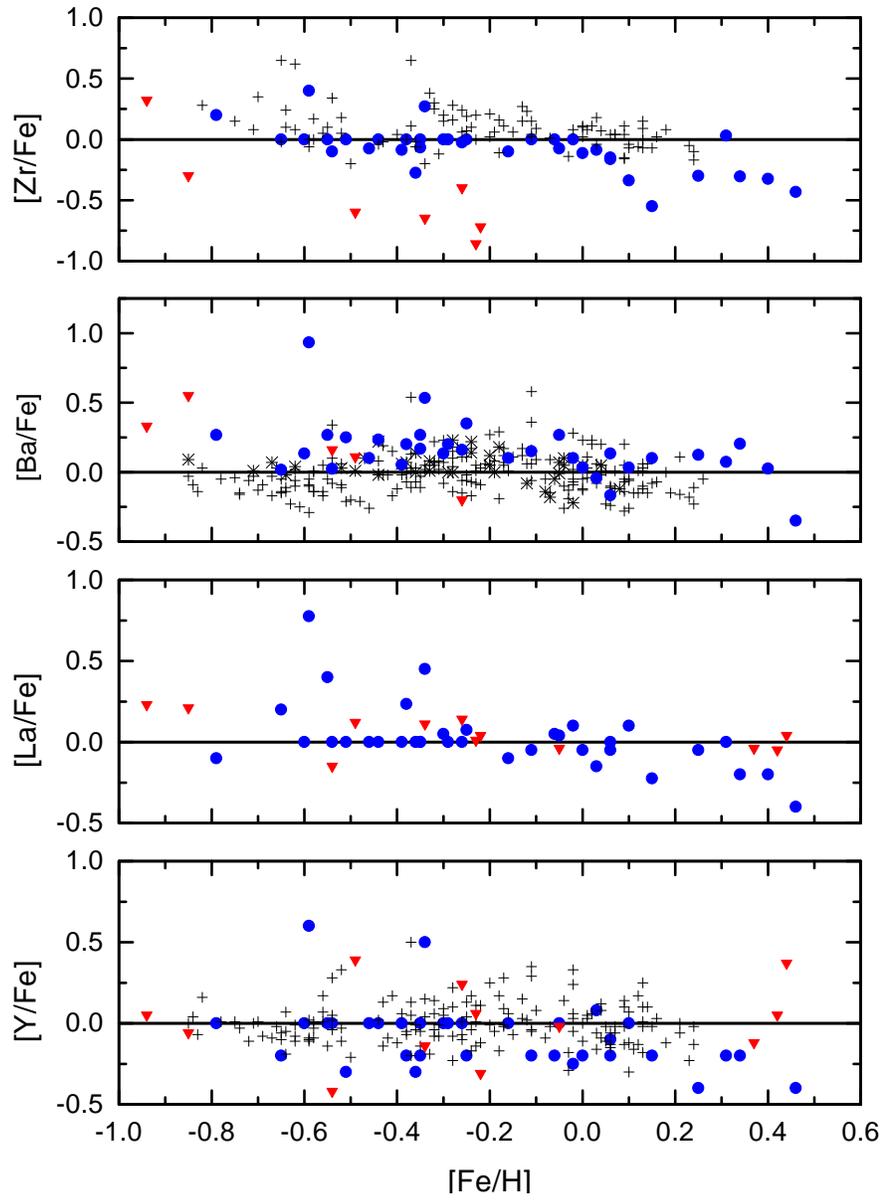}
\caption{[s-elements/Fe] vs. [Fe/H] relation for Zr, Ba, La and Y. Symbols as in Figure 5.}
\end{figure}

\placefigure{fg8}

\section{Discussion}

We draw a series of  inferences regarding the sample stars:

1. Abundance trends depend on the star formation history of a given population. In a star forming region,
the yields of the SNe II are the first to enrich the interestellar medium (ISM). A large fraction of the
iron peak elements is produced later by the SNe Ia, after
 $\sim$ 1-2 Gyr (e.g. Yoshii et al. 1996).
Therefore, in the [$\alpha$/Fe] vs. [Fe/H] plot, a constant [$\alpha$/Fe] ratio or a  "plateau" is seen
at low metallicities. When the first SNe Ia begin to explode, this ratio has a downturn, and a knee
in the [$\alpha$/Fe] vs. [Fe/H] curve is seen (see e.g. Figure 1 of RM00,  McWilliam 1997, Barbuy 1988 and
Barbuy \& Erdelyi-Mendes 1989).

As well illustrated in Fig. 4 of Matteucci \& Brocato (1990, hereafter MB90),
for the bulge, the point of downturn of [$\alpha$/Fe] vs. [Fe/H]
 curves at higher metallicities is a result of a
star formation rate more efficient than in the solar vicinity
combined to a  timescale of formation of the system much shorter (estimated to be of the order 10$^7$ yr
for the bulge by MB90, but probably somewhat less extreme for the present sample).
For the present sample the point of inflexion for [O/Fe], [Mg/Fe] and also that of the $r$-element [Eu/Fe] is
around [Fe/H] $\approx$ -0.2, only somewhat less than the estimations for the bulge by MB90.
In other words,
the knee position in the plot gives information on the star formation rate (SFR).
If the SFR is high, a higher metallicity is achieved before the first SNe Ia explode, and the knee is
deviated to the right. A fast
 SF has certainly occurred for our sample stars as well as for the bulge.

\smallskip

 2. The level of overabundance of the [$\alpha$/Fe] plateau at
 low metallicities gives hints on the initial
mass function (IMF) of the system. The larger the number of
 massive stars, the highest the plateau.

\smallskip

3. The slow decreasing trends of Si and Ca with [Fe/H] support the idea that such elements are produced by
both SNe II and SNe Ia (Matteucci et al. 1999). Ti follows the same behavior of  Si and Ca, also hinting
for a production by both SNe types as discussed in previous work (McWilliam 1995a,b). Nevertheless,
Ti still has a poorly known formation process  and yield predictions (Timmes et al. 1995; Nagataki 1999;
Thielemann et al. 2002). Moreover, although in the bulge stars of MR94 and RM00
the other [$\alpha$/Fe]-elements show a similar pattern to our sample, Ti has a puzzling behavior,
with an apparently constant value for almost all the present metallicity range.

4. An interesting feature of the $\alpha$-elements distribution in the present sample is the point where
the abundance ratios stop the decreasing trend. 
This occurs for all $\alpha$-elements, with many subsolar values for
roughly the same metallicity, [Fe/H] $\sim$ -0.1 dex. Nevertheless, for Ti the subsolar ratios can  
indicate a systematic error or NLTE effect as discussed by Luck \& Bond (1985) and Prochaska et al. (2000).
If real, such behavior could be explained by a starburst formation followed by a long interval
with negligible SF (Gilmore \& Wyse 1991). During this
interval the SNe Ia proceed exploding, raising the metallicity of the ISM and eventually decreasing the
[$\alpha$/Fe] ratios of the newly forming stars. 

\smallskip

5. $s$-elements indicate a late evolution of the population given that they are mainly produced by AGB
stars in typical timescales of 2-3 Gyr (e.g. Pagel \& Tautvai\v{s}iene 1997).
However a small contribution by the $r$-process is also expected (Truran 1981). A useful diagnostic
of the relative importance of the neutron capture processes is the [Eu/$s$] ratio, given that Eu is an almost
pure (95\%) $r$-process element. In Figure 9 the [Eu/$s$] vs. [Fe/H] relation for
Ba, La and Y are plotted, as well as for the $\alpha$-element Mg. [Eu/s-element] curves show
a very slow decrease with metallicity until [Fe/H] $\sim$ -0.10, possibly related to the delayed enrichment of 
the interstellar medium in $s$-elements by AGB stars relative to SNe II. Such decrease is in contrast with the almost 
constant behavior of [Eu/Mg] over all the metallicity range, suggesting the common origin of these elements. For
higher metallicity stars, an increasing trend is seen also hinting for a late burst of SF.  

\begin{figure}
\epsscale{0.6}
\label{fg9}
\plotone{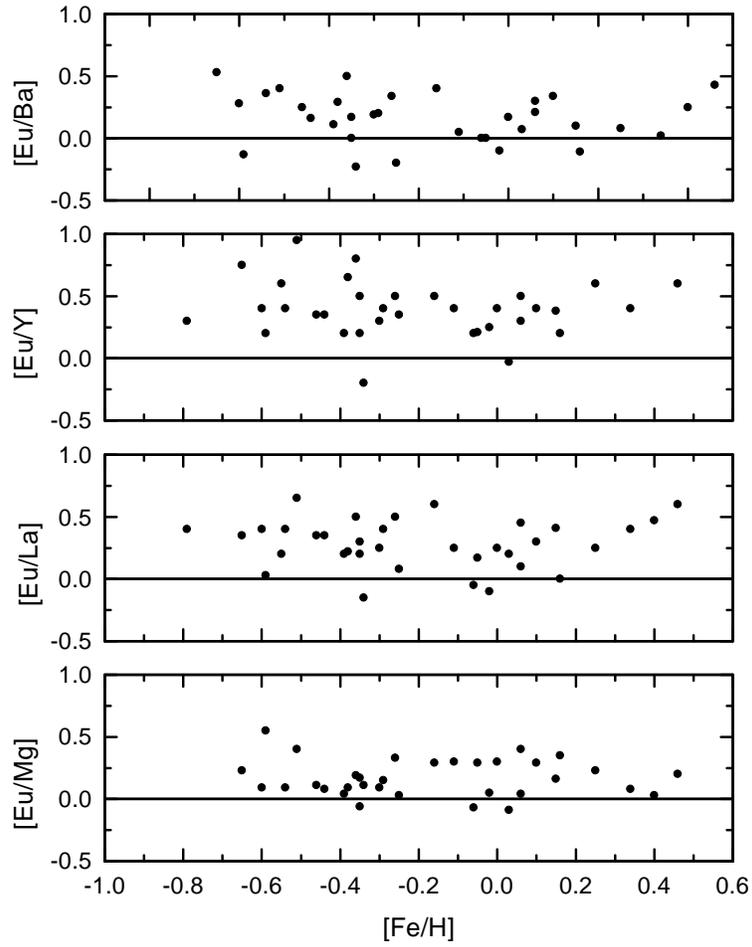}
\caption{[Eu/X] vs. [Fe/H] relation for the $s$-elements Ba, La and Y, and for the
$\alpha$-element Mg.}
\end{figure}

\placefigure{fg9}

\smallskip

6.  The high metallicity of the stars and high eccentricity of the orbits of the present sample
 points to a probable inner disk or bulge origin. The old and narrow age
range of the sample (Grenon 1998) also support this hypothesis. The presence of inner Galaxy stars
in the solar neighborhood has
been predicted by dynamical simulations of particles under a bar potential well, which triggers chaotic
"hot" movements towards the center and anti-center directions (e.g. Zhao et al. 1994; Raboud et al.
1998; Fux 1997, 2001). From the
 chemical distribution of our sample, a fast star formation within
a possible starburst regime is inferred. If such stars are indeed bulge stars,
this result is consistent with an inside-out scenario of Galaxy formation, with a short timescale
for bulge formation (MB90; Wyse \& Gilmore 1992; Matteucci \& Romano 1999,
Matteucci et al. 1999). The inner disk origin hypothesis is more difficult to be evaluated
because of the paucity of data for the metallicity distribution, age, and chemical features of stars of this
region.

\section{Summary}

In the present work we show that on chemical and kinematical grounds the bulgelike sample
represents a distinct population when compared to the disk population.
Eu/Fe, O/Fe and Mg/Fe abundance ratios
show an enhanced pattern relative to disk stars, a signature of a higher enrichment by SNe II yields,
indicating a faster evolution process when compared to the disk.

The point of inflexion of [O, Mg, Eu/Fe] vs. [Fe/H]
 curves at metallicities [Fe/H] $\approx$ -0.2
are a result of a star formation rate more efficient than in the solar vicinity, combined to a shorter
timescale of formation of the system. The point of 
inflexion  derived from our observations is
at a somewhat lower metallicity than the estimations for the bulge by MB90.

Si and Ca, on the other hand, follow the disk and bulge trends, consistent with the theories of different
sites of nucleosynthetic formation. Ti has a puzzling behavior when compared to bulge stars. In our sample
it follows the disk trend while in bulge stars it has an almost constant and high value.

The differences between the $\alpha$-elements O, Mg on one hand,
and Si and Ca on the other, has already been found in other samples,
in particular the field bulge stars studied by MR94 and RM00,
as well as in bulge clusters by Barbuy et al. (1999).

From the chemical distributions of our sample, a fast star formation
within a possible starburst regime is inferred. If such stars are indeed bulge stars,
this result is consistent with an inside-out scenario of Galaxy formation,
with a short timescale
for bulge formation (MB90; Wyse \& Gilmore 1992; Matteucci \& Romano
1999, Matteucci et al. 1999).

\bigskip
\bigskip

$Acknowledgments$  L. P. acknowledges R. Gallino, K. Nomoto, G. Tautvai\v{s}iene, T. Tsujimoto, Y.
Z. Qian and C. Fryer for very useful discussions during the "Stellar Abundances and Nucleosynthesis
Conference", and the organizers of this excellent meeting, in particular G. Wallerstein, for this 
opportunity and financial support. We gratefully thank A. McWilliam and V. Hill for very useful 
discussions on hyperfine structures. L. P. thanks A. Ardila-Rodriguez and R. R. Rosa for technical support.    
L. P. acknowledges FAPESP PhD and pos-doc fellowships n$^{\rm o}$ 98/00014-0 and 01/14594-2. We 
acknowledge FAPESP project n$^{\rm o}$ 1998/10138-8.



\begin{deluxetable}{cccccccccc}
\label{tb1}
\tablecolumns{10}
\tablewidth{0pc}
\tablecaption{Atmospheric Parameters for the sample stars}
\tablewidth{0pt}
\tablehead{
\colhead{Name} & \colhead{Number} & \colhead{T$_{\rm eff}$} & \colhead{log g} & \colhead{[FeI/H]}
& \colhead{$\sigma$(FeI)} & \colhead{[FeII/H]} & \colhead{$\sigma$(FeII)}
& \colhead{$\xi(\rm kms^{-1})$}
}
\startdata
HD 143016  &  b1  &  5575  &  4.11  & -0.59  &  0.12  &  -0.35  &  0.14  &  0.70 \\
HD 143102  &  b2  &  5500  &  3.85  & -0.02  &  0.13  &  0.03  &  0.16  &  1.05  \\
HD 148530  &  b3  &  5350  &  4.43  & -0.12  &  0.14  &  0.10  &  0.15  &  0.80 \\
HD 149256  &  b4  &  5350  &  3.73  & 0.30  &  0.16  &  0.34  &  0.16  &  0.80 \\
HD 152391  &  b5  &  5300  &  4.45  & -0.12  &  0.14  &  -0.05  &  0.12  &  0.50 \\
HD 326583  &  b6  &  5600  &  3.81  & -0.44  &  0.13  &  -0.30  &  0.17  &  0.60 \\
HD 175617  &  b7  &  5550  &  4.56  & -0.50  &  0.09  &  -0.44  &  0.14  &  0.60 \\
HD 178737  &  b8  &  5575  &  3.90  & -0.37  &  0.13  &  -0.35  &  0.14  &  0.60  \\
HD 179764  &  b9  &  5450  &  4.26  & -0.12  &  0.13  &  0.06  &  0.10  &  1.10 \\
HD 181234  &  b10 &  5350  &  4.25  &  0.25  &  0.14  &  0.40  &  0.19  &  1.10 \\
HD 184846  &  b11 &  5600  &  4.40  &  -0.29  &  0.10  &  0.06  &  0.07  &  0.50 \\
BD-176035  &  b12  &  4750  &  4.36  & 0.21  &  0.15  &  0.46  &  0.13  &  0.70   \\
HD 198245  &  b13  &  5650  &  4.31  & -0.69  &  0.11  &  -0.60  &  0.10  &  0.60  \\
HD 201237  &  b14  &  4950  &  4.08  & 0.13  &  0.16  &  0.15  &  0.18  &  0.50  \\
HD 211276  &  b15  &  5500  &  4.05  & -0.61  &  0.13  &  -0.39  &  0.14  &  0.50 \\
HD 211532  &  b16  &  5350  &  4.46  & -0.69  &  0.15  &  -0.54  &  0.18  &  0.80 \\
HD 211706  &  b17  &  5800  &  4.25  & -0.04  &  0.12  &  0.16  &  0.22  &  0.80  \\
HD 214059  &  b18  &  5550  &  3.81  & -0.17  &  0.11  &  -0.38  &  0.14  &  0.75  \\
CD-40 15036 & b19  &  5350 &  4.34  & -0.11  &  0.12  &  0.00  &  0.13   &  0.70 \\
HD 219180  &  b20  &  5400  &  4.35  & -0.73  &  0.15  &  -0.46  &  0.03  &  0.65 \\
HD 220536  &  b21  &  5850  &  4.17  & -0.23  &  0.11  &  -0.11  &  0.10  &  0.50 \\
HD 220993  &  b22  &  5600  &  4.15  & -0.33  &  0.12  &  -0.16  &  0.08  &  0.80 \\
HD 224383  &  b23  &  5800  &  4.14  & -0.07  &  0.12  &  -0.06  &  0.11  &  1.00 \\
HD   4308  &  b24  &  5600  &  4.31  & -0.45  &  0.10  &  -0.26  &  0.09  &  0.80 \\
HD   6734  &  b25  &  5000  &  3.40  & -0.50  &  0.09  &  -0.36  &  0.10  &  0.75 \\
HD   8638  &  b26  &  5500  &  4.38  & -0.47  &  0.10  &  -0.29  &  0.13  &  0.60 \\
HD   9424  &  b27  &  5350  &  4.35  & -0.03  &  0.14  &  0.25  &  0.12  &  0.70 \\
HD  10576  &  b28  &  5850  &  4.00  & -0.16  &  0.08  &  -0.02  &  0.10  &  1.25 \\
HD  10785  &  b29  &  5850  &  4.16  & -0.25  &  0.11  &  -0.25  &  0.16  &  1.20 \\
HD  11306  &  b30  &  5200  &  4.09  & -0.63  &  0.11  &  -0.98  &  0.19  &  0.80 \\
HD  11397  &  b31  &  5400  &  4.34  & -0.76  &  0.10  &  -0.59  &  0.15  &  0.65 \\
HD  14282  &  b32  &  5800  &  3.91  & -0.43  &  0.12  &  -0.34  &  0.14  &  0.65 \\
HD  16623  &  b33  &  5700  &  4.26  & -0.60  &  0.09  &  -0.51  &  0.15  &  0.60 \\
BD-02 603  &  b34  &  5450  &  3.75  & -0.77  &  0.12  &  -0.79  &  0.14  &  0.70 \\
HD  21543  &  b35  &  5650  &  4.37  & -0.62  &  0.10  &  -0.55  &  0.15  &  0.50 \\
\enddata
\end{deluxetable}

\begin{deluxetable}{ccccc}
\label{tb2}
\tabletypesize{\scriptsize}
\tablecolumns{5}
\tablewidth{0pc}
\tablecaption{Atomic Line List}
\tablewidth{0pt}
\tablehead{
\colhead{Element} & \colhead{$\lambda$} & \colhead{$\xi$$_{exc}$} &
\colhead{log $gf$} & \colhead{Reference}
}
\startdata
\ion{Ca}{1}  &  6161.30  &  2.52  &  \llap{$-$}1.03  &  S  \\
\ion{Ca}{1}  &  6455.60  &  2.52  &  \llap{$-$}1.28  &  B  \\
\ion{Ca}{1}  &  6508.85  &  2.52  &  \llap{$-$}2.50  &  MR  \\
\ion{Ca}{1}  &  6572.79  &  0.00  &  \llap{$-$}4.30  &  N  \\
 \ion{O}{1}   &  6300.31  &  0.00  &  \llap{$-$}9.76  &  C  \\
\ion{Ti}{1}  &  6303.77  &  1.4   &  \llap{$-$}1.57  &  N  \\
\ion{Ti}{1}  &  6312.24  &  1.46  &  \llap{$-$}1.55  &  N  \\
\ion{Ti}{1}  &  6336.11  &  1.44  &  \llap{$-$}1.74  &  N  \\
\ion{Ti}{1}  &  6554.22  &  1.44  &  \llap{$-$}1.22  &  N  \\
\ion{Ti}{1}  &  6599.13  &  0.9   &  \llap{$-$}2.09  &  N  \\
\ion{Ti}{1}  &  6743.13  &  0.9   &  \llap{$-$}1.63  &  N  \\
\ion{La}{2}  &  6390.48  &  0.320 &  \llap{$-$}1.52  &  MR  \\
\ion{La}{2}  &  6774.26  &  0.130 &  \llap{$-$}1.81  &  MR  \\
\ion{Eu}{2}  &  6645.13  &  1.38  &            0.20  &  H  \\
\ion{Mg}{1}  &  6319.24  &  5.110 &  \llap{$-$}2.22  &  MR  \\
\ion{Mg}{1}  &  7387.70  &  5.150 &  \llap{$-$}1.15  &  B  \\
\ion{Si}{1}  &  6237.33  &  5.610 &  \llap{$-$}1.01  &  MR  \\
\ion{Si}{1}  &  6721.84  &  5.860 &  \llap{$-$}1.17 &  MR  \\
\ion{Si}{1}  &  7226.21  &  5.610 &  \llap{$-$}1.44 &  MR  \\
\ion{Si}{1}  &  7405.79  &  5.610 &  \llap{$-$}0.63 &  B  \\
\ion{Si}{1}  &  7415.96  &  5.610 &  \llap{$-$}0.73 &  MR  \\
\ion{Y}{2}   &  6795.41  &  1.73  &  \llap{$-$}1.25 &  MR  \\
\ion{Zr}{2}  &  6114.85  & 1.67   &  \llap{$-$}1.7  &  S  \\
\ion{Zr}{1}  &  6134.57  &  0.00  &  \llap{$-$}1.28 &  S  \\
\ion{Zr}{1}  &  6140.46  &  0.52  &  \llap{$-$}1.41 &  S  \\
\ion{Zr}{1}  &  6762.40  &  0.00  &  \llap{$-$}1.3  &  B    \\
\ion{Ba}{2}  &  4554.03  &  0.00  &            0.17 &  F \\
\ion{Ba}{2}  &  4934.10  &  0.00  &  \llap{$-$}1.15 &  F*  \\
\ion{Ba}{2}  &  6496.91  &  0.60  &  \llap{$-$}0.38 &  F  \\ \enddata
\tablerefs{N - NIST (http:/\/\ phy\-sics\-.nist\-.gov\-/cgi\--bin\-/AtData\-/main-asd),
MR - McWilliam \& Rich (1994), S - Smith et al. (2000), C - Castro et al. (1997), B - 
Barbuy et al. (1999, 2002), F - Fran\c cois (1996), F* - Fran\c cois (private communication),  
Hill et al. (2000).}
\end{deluxetable}

\clearpage

\begin{deluxetable}{ccccccccccc}
\label{tb3}
\tablecolumns{11}
\tablewidth{0pc}
\tablecaption{Abundances} \tablewidth{0pt}
\tabletypesize{\footnotesize} \tablehead{ \colhead{Star} &
\colhead{[Ca/Fe]} & \colhead{[Si/Fe]} & \colhead{[Ti/Fe]} &
\colhead{[Mg/Fe]} & \colhead{[Ba/Fe]} & \colhead{[Zr/Fe]} &
\colhead{[La/Fe]} & \colhead{[Y/Fe]} & \colhead{[Eu/Fe]} &
\colhead{[O/Fe]} }
\startdata	
b1	&	\llap{$-$}0.11	$\pm$	0.12	&	0.08	$\pm$	0.06	&	\llap{$-$}0.07	$\pm$	0.10	&	0.14	$\pm$	0.01	&	0.13	$\pm$	0.15	&	\llap{$<$}-0.07	$\pm$	0.06	&	0.00	$\pm$	0.00	&	\llap{$<$}-0.20	&	0.30	&	0.44	\\
b2	&	\llap{$-$}0.02	$\pm$	0.04	&	0.05	$\pm$	0.04	&	\llap{$-$}0.04	$\pm$	0.09	&	0.14	$\pm$	0.06	&	\llap{$-$}0.02	$\pm$	0.18	&	\llap{$<$}-0.09	$\pm$	0.12	&	\llap{$-$}0.15	$\pm$	0.21	&	0.08	&	0.05	&	\llap{$<$} -0.2	\\
b3	&	\llap{$-$}0.09	$\pm$	0.07	&	0.03	$\pm$	0.12	&	\llap{$-$}0.10	$\pm$	0.06	&	0.12	$\pm$	0.12	&	0.06	$\pm$	0.15	&	\llap{$<$}-0.34	$\pm$	0.05	&	0.10	$\pm$	0.14	&	\llap{$<$}0.00	&	0.40	&	\llap{$-$}0.1	\\
b4	&	0.01	$\pm$	0.10	&	0.15	$\pm$	0.04	&	\llap{$-$}0.13	$\pm$	0.04	&	0.13	$\pm$	0.11	&	0.18	$\pm$	0.05	&	\llap{$<$}-0.31	$\pm$	0.16	&	\llap{$-$}0.20	$\pm$	0.14	&	\llap{$-$}0.20	&	0.20	&	\llap{$-$}0.14	\\
b5	&	\llap{$-$}0.17	$\pm$	0.10	&	\llap{$-$}0.03	$\pm$	0.05	&	\llap{$-$}0.24	$\pm$	0.09	&	\llap{$-$}0.08	$\pm$	0.04	&	0.21	$\pm$	0.21	&	\llap{$<$}-0.08	$\pm$	0.15	&	\llap{$<$}0.04	$\pm$	0.20	&	\llap{$<$}0.00	&	0.21	&	\llap{$<$} 0.13	\\
b6	&	0.01	$\pm$	0.13	&	0.03	$\pm$	0.08	&	0.06	$\pm$	0.12	&	0.22	$\pm$	0.05	&	0.11	$\pm$	0.16	&	\llap{$<$}0.00	$\pm$	0.00	&	0.05	$\pm$	0.07	&	0.00	&	0.30	&	0.22	\\
b7	&	0.17	$\pm$	0.12	&	0.11	$\pm$	0.09	&	0.27	$\pm$	0.15	&	0.27	$\pm$	0.04	&	0.19	$\pm$	0.08	&	\llap{$<$}0.00	$\pm$	0.00	&	\llap{$<$}0.00	$\pm$	0.00	&	0.00	&	0.35	&	0.53	\\
b8	&	0.06	$\pm$	0.06	&	0.15	$\pm$	0.06	&	0.11	$\pm$	0.08	&	0.26	$\pm$	0.01	&	0.20	$\pm$	0.03	&	\llap{$<$}0.00	$\pm$	0.00	&	\llap{$<$}0.00	$\pm$	0.00	&	\llap{$<$}0.00	&	0.20	&	0.31	\\
b9	&	\llap{$-$}0.04	$\pm$	0.06	&	\llap{$-$}0.01	$\pm$	0.07	&	\llap{$-$}0.05	$\pm$	0.05	&	0.07	$\pm$	0.05	&	\llap{$-$}0.11	$\pm$	0.15	&	\llap{$<$}-0.16	$\pm$	0.13	&	\llap{$<$}0.00	$\pm$	0.00	&	\llap{$<$}-0.20	&	0.10	&	0.05	\\
b10	&	0.00	$\pm$	0.03	&	0.06	$\pm$	0.08	&	\llap{$-$}0.01	$\pm$	0.02	&	0.24	$\pm$	0.01	&	0.02	$\pm$	0.18	&	\llap{$<$}-0.33	$\pm$	0.05	&	\llap{$<$}-0.20	$\pm$	0.14	&	  -	    &	0.27	&	\llap{$-$}0.22	\\
b11	&	\llap{$-$}0.17	$\pm$	0.11	&	\llap{$-$}0.07	$\pm$	0.03	&	\llap{$-$}0.23	$\pm$	0.09	&	0.00	$\pm$	0.00	&	0.10	$\pm$	0.12	&	\llap{$<$}-0.10	$\pm$	0.17	&	\llap{$<$}-0.05	$\pm$	0.07	&	\llap{$-$}0.10	&	0.40	&	0.3	\\
b12	&	\llap{$-$}0.27	$\pm$	0.11	&	\llap{$-$}0.07	$\pm$	0.11	&	\llap{$-$}0.29	$\pm$	0.20	&	0.00	$\pm$	0.00	&	\llap{$-$}0.23	$\pm$	0.35	&	\llap{$<$}-0.32	$\pm$	0.07	&	\llap{$<$}-0.40	$\pm$	0.00	&	\llap{$<$}-0.40	&	0.20	&	-	\\
b13	&	0.08	$\pm$	0.10	&	0.05	$\pm$	0.10	&	0.14	$\pm$	0.14	&	0.32	$\pm$	0.09	&	0.12	$\pm$	0.06	&	\llap{$<$}0.01	$\pm$	0.00	&	\llap{$<$}0.00	$\pm$	0.00	&	\llap{$<$}0.00	&	0.40	&	\llap{$<$} 0.1	\\
b14	&	0.23	$\pm$	0.15	&	\llap{$-$}0.21	$\pm$	0.09	&	0.19	$\pm$	0.12	&	0.03	$\pm$	0.04	&	0.08	$\pm$	0.06	&	\llap{$<$}-0.43	$\pm$	0.25	&	\llap{$-$}0.23	$\pm$	0.04	&	\llap{$-$}0.20	&	0.18	&	\llap{$<$} -0.25	\\
b15	&	\llap{$-$}0.08	$\pm$	0.08	&	\llap{$-$}0.16	$\pm$	0.17	&	\llap{$-$}0.03	$\pm$	0.12	&	0.17	$\pm$	0.21	&	0.09	$\pm$	0.09	&	\llap{$<$}-0.05	$\pm$	0.10	&	\llap{$<$}0.00	$\pm$	0.00	&	\llap{$<$}0.00	&	\llap{$<$}0.20	&	-	\\
b16	&	0.11	$\pm$	0.20	&	\llap{$-$}0.03	$\pm$	0.15	&	0.27	$\pm$	0.09	&	0.31	$\pm$	0.10	&	0.04	$\pm$	0.13	&	\llap{$<$}-0.04	$\pm$	0.17	&	\llap{$<$}0.00	$\pm$	0.00	&	0.00	&	0.40	&	\llap{$<$} 0.15	\\
b17	&	\llap{$-$}0.31	$\pm$	0.09	&	\llap{$-$}0.14	$\pm$	0.06	&	\llap{$-$}0.36	$\pm$	0.11	&	\llap{$-$}0.35	$\pm$	0.21	&	0.11	$\pm$	0.13	&	\llap{$<$}0.05	$\pm$	0.06	&	\llap{$<$}0.00	$\pm$	0.00	&	\llap{$-$}0.20	&	\llap{$<$}0.00	&	0.14	\\
b18	&	0.10	$\pm$	0.11	&	0.19	$\pm$	0.09	&	0.12	$\pm$	0.06	&	0.36	$\pm$	0.06	&	0.16	$\pm$	0.05	&	\llap{$<$}0.01	$\pm$	0.00	&	0.24	$\pm$	0.16	&	\llap{$-$}0.20	&	 -  	&	-	\\
b19	&	\llap{$-$}0.07	$\pm$	0.08	&	\llap{$-$}0.13	$\pm$	0.06	&	\llap{$-$}0.12	$\pm$	0.07	&	\llap{$-$}0.10	$\pm$	0.14	&	0.03	$\pm$	0.06	&	\llap{$<$}-0.11	$\pm$	0.09	&	\llap{$<$}-0.05	$\pm$	0.07	&	\llap{$<$}-0.20	&	0.20	&	0.05	\\
b20	&	\llap{$-$}0.06	$\pm$	0.10	&	0.07	$\pm$	0.14	&	0.01	$\pm$	0.10	&	0.24	$\pm$	0.17	&	0.10	$\pm$	0.10	&	\llap{$<$}-0.08	$\pm$	0.15	&	\llap{$<$}0.00	$\pm$	0.00	&	\llap{$<$}0.00	&	0.35	&	-	\\
b21	&	\llap{$-$}0.30	$\pm$	0.07	&	\llap{$-$}0.18	$\pm$	0.05	&	\llap{$-$}0.16	$\pm$	0.14	&	\llap{$-$}0.10	$\pm$	0.00	&	0.15	$\pm$	0.17	&	\llap{$<$}0.00	$\pm$	0.00	&	\llap{$<$}-0.05	$\pm$	0.07	&	\llap{$-$}0.20	&	0.20	&	\llap{$-$}0.1	\\
b22	&	\llap{$-$}0.09	$\pm$	0.09	&	0.01	$\pm$	0.08	&	\llap{$-$}0.05	$\pm$	0.08	&	0.21	$\pm$	0.13	&	0.10	$\pm$	0.10	&	\llap{$<$}-0.10	$\pm$	0.12	&	\llap{$<$}-0.10	$\pm$	0.00	&	\llap{$<$}0.00	&	0.50	&	\llap{$<$} 0.1	\\
b23	&	\llap{$-$}0.06	$\pm$	0.12	&	\llap{$-$}0.04	$\pm$	0.06	&	\llap{$-$}0.12	$\pm$	0.06	&	0.07	$\pm$	0.02	&	0.00	$\pm$	0.20	&	\llap{$<$}0.00	$\pm$	0.00	&	0.05	$\pm$	0.07	&	\llap{$<$}-0.20	&	0.00	&	0.17	\\
b24	&	\llap{$-$}0.08	$\pm$	0.07	&	0.05	$\pm$	0.06	&	\llap{$-$}0.05	$\pm$	0.08	&	0.17	$\pm$	0.01	&	0.16	$\pm$	0.05	&	\llap{$<$}-0.03	$\pm$	0.05	&	\llap{$<$}0.00	$\pm$	0.00	&	\llap{$<$}0.00	&	0.50	&	\llap{$<$} 0.2	\\
b25	&	0.05	$\pm$	0.07	&	0.12	$\pm$	0.03	&	0.13	$\pm$	0.04	&	0.32	$\pm$	0.05	&	0.00	$\pm$	0.00	&	\llap{$<$}-0.28	$\pm$	0.15	&	0.00	$\pm$	0.00	&	\llap{$<$}-0.30	&	0.50	&	0.25	\\
b26	&	0.00	$\pm$	0.11	&	0.06	$\pm$	0.05	&	0.04	$\pm$	0.08	&	0.25	$\pm$	0.00	&	0.20	$\pm$	0.00	&	\llap{$<$}0.00	$\pm$	0.00	&	\llap{$<$}0.00	$\pm$	0.00	&	\llap{$<$}0.00	&	0.40	&	0.53	\\
b27	&	\llap{$-$}0.20	$\pm$	0.08	&	\llap{$-$}0.08	$\pm$	0.06	&	\llap{$-$}0.26	$\pm$	0.08	&	\llap{$-$}0.03	$\pm$	0.11	&	0.12	$\pm$	0.15	&	\llap{$<$}-0.30	$\pm$	0.16	&	\llap{$-$}0.05	$\pm$	0.07	&	\llap{$<$}-0.40	&	\llap{$<$}0.20	&	\llap{$-$}0.1	\\
b28	&	\llap{$-$}0.20	$\pm$	0.12	&	\llap{$-$}0.13	$\pm$	0.07	&	\llap{$-$}0.20	$\pm$	0.05	&	\llap{$-$}0.05	$\pm$	0.00	&	0.10	$\pm$	0.11	&	\llap{$<$}0.00	$\pm$	0.00	&	0.10	$\pm$	0.07	&	\llap{$-$}0.25	&	0.00	&	\llap{$<$}0.1	\\
b29	&	0.00	$\pm$	0.18	&	0.07	$\pm$	0.05	&	0.03	$\pm$	0.12	&	0.13	$\pm$	0.04	&	0.35	$\pm$	0.09	&	\llap{$<$}0.00	$\pm$	0.00	&	0.08	$\pm$	0.11	&	\llap{$<$}-0.20	&	0.15	&	-	\\
b30	&	0.15	$\pm$	0.12	&	0.09	$\pm$	0.09	&	0.26	$\pm$	0.08	&	0.33	$\pm$	0.04	&	0.02	$\pm$	0.04	&	\llap{$<$}0.00	$\pm$	0.00	&	\llap{$<$}0.20	$\pm$	0.14	&	\llap{$<$}-0.20	&	0.55	&	\llap{$<$}0.1	\\
b31	&	0.00	$\pm$	0.09	&	0.19	$\pm$	0.02	&	\llap{$-$}0.02	$\pm$	0.06	&	0.26	$\pm$	0.04	&	0.93	$\pm$	0.12	&	\llap{$<$}0.40	$\pm$	0.27	&	0.78	$\pm$	0.11	&	0.60	&	0.80	&	\llap{$<$}0.2	\\
b32	&	\llap{$-$}0.02	$\pm$	0.13	&	0.07	$\pm$	0.02	&	\llap{$-$}0.12	$\pm$	0.10	&	0.19	$\pm$	0.01	&	0.53	$\pm$	0.15	&	\llap{$<$}0.27	$\pm$	0.23	&	0.45	$\pm$	0.07	&	0.50	&	0.30	&	\llap{$<$}0.28	\\
b33	&	\llap{$-$}0.02	$\pm$	0.08	&	0.10	$\pm$	0.04	&	0.00	$\pm$	0.08	&	0.25	$\pm$	0.18	&	0.25	$\pm$	0.00	&	\llap{$<$}0.00	$\pm$	0.00	&	\llap{$<$}0.00	$\pm$	0.00	&	\llap{$<$}-0.30	&	0.65	&	0.4	\\
b34	&	0.31	$\pm$	0.10	&	0.17	$\pm$	0.09	&	0.27	$\pm$	0.06	&	0.48	$\pm$	0.00	&	0.27	$\pm$	0.12	&	\llap{$<$}0.20	$\pm$	0.23	&	\llap{$<$}-0.10	$\pm$	0.14	&	\llap{$<$}0.00	&	0.30	&	-	\\
b35	&	0.09	$\pm$	0.07	&	0.21	$\pm$	0.06	&	0.17	$\pm$	0.15	&	0.25	$\pm$	0.18	&	0.27	$\pm$	0.13	&	\llap{$<$}0.00	$\pm$	0.00	&	\llap{$<$}0.40	$\pm$	0.28	&	\llap{$<$}0.00	&	0.60	&	0.51	\\
\enddata
\end{deluxetable}

\clearpage

\begin{deluxetable}{crrrrrrrrrrrrrrrrrrr}
\label{tb4}
\tabletypesize{\scriptsize}
\tablecolumns{14}
\tablecaption{Measured equivalent widths for the atomic lines: b1 to b17}
\tablewidth{0pt}
\tablehead{
\colhead{Element} & \colhead{$\lambda$} & \colhead{b1} & \colhead{b2} & \colhead{b3}  & \colhead{b4} & \colhead{b5} 
& \colhead{b6} & \colhead{b7} & \colhead{b8} & \colhead{b9} &
\colhead{b10} & \colhead{b11} & \colhead{b12} & \colhead{b13} &  \colhead{b14} & \colhead{b15}
& \colhead{b16} & \colhead{b17} 
}
\startdata
Ba II   &  4554.03 & 155 & 154 & 186 & 202 & 202 & 150 & 152 & 159 & 177 & 194 & 158 & 212 & 134 & 208 & 149 & 145& 167 \\
Ba II   &  4934.10 & 159 & 172 & 164 & 176 & 181 & 160 & 162 & 157 & 172 & 146 & 159 & 182 & 140 & 187 & 146 & 147& 163 \\
Ba II   &  6496.93 &  88 & 116 & 104 & 123 & 124 &  93 &  81 &  92 & 105 & 111 &  88 & 118 &  72 & 106 &  89 & 58 & 106 \\
Zr II   &  6114.85 &   - &   - &   - &   - &   - &   - &   - &   - &   - &  11 &   - &   - &   - &   - &   - &  - &   - \\
Zr I    &  6134.57 &   - &   3 &   5 &   9 &   5 &   - &   2 &   - &   3 &   - &   - &  29 &   - &  19 &   - & 17 &   - \\
Zr I    &  6140.46 &   - &   - &   - &   - &   - &   - &   - &   - &   - & 141 &   - &   5 &   - &   5 &   - &  7 &   - \\
Zr I    &  6762.40 &   - &   - &   - &  14 &   5 &   - &   - &   - &   3 &   - &   - &  15 &   3 &  11 &   - &  - &   - \\
La II   &  6390.48 &   - &   5 &   - &   7 &   - &   8 &   - &   - &   6 &  10 &   - &   9 &   - &   6 &   - &  - &   - \\
La II   &  6774.26 &   4 &   - &   - &   - &   - &   - &   2 &   - &   - &  10 &   - &   - &   - &   5 &   - &  - &   - \\
Y  II   &  6795.41 &   - &  10 &   - &   6 &   6 &   - &   5 &   - &   - &   - &   - &   - &   - &   3 &   - &  - &   - \\
Eu II   &  6645.10 &   7 &  13 &   6 &  15 &   7 &   - &   5 &  14 &   8 &  12 &  22 &  11 &   6 &   9 &   - & 27 &   9 \\
Ca I    &  6161.30 &  33 &  85 &  89 & 113 &  64 &  61 &  61 &  60 &  88 & 108 &  71 &   - &  50 &   - &   - & 62 &  65 \\
Ca I    &  6455.60 &  45 &  72 &  77 &  88 &  72 &  52 &  56 &  52 &  76 &  95 &  62 & 113 &  40 & 104 &  45 &  - &  65 \\
Ca I    &  6508.85 &   - &  19 &  17 &  35 &  15 &   7 &   7 &  10 &  17 &  37 &  13 &  44 &   3 &  44 &  10 & 11 &   8 \\
Ca I    &  6572.79 &  20 &  60 &  65 &  84 &  57 &  38 &  45 &  34 &  69 &  85 &  37 & 127 &  23 & 116 &  25 & 44 &  28 \\
Si I    &  6237.33 &  47 &  85 &  64 &  95 &  57 &  57 &  42 &  55 &  68 &  87 &  58 &  55 &  38 &  47 &  27 & 32 &  69 \\
Si I    &  6721.84 &  31 &  64 &  52 &  84 &  45 &  41 &  29 &  41 &  54 &  81 &  45 &  58 &  26 &  41 &  24 & 26 &  51 \\
Si I    &  7226.21 &  34 &  56 &  46 &  66 &  44 &  35 &  25 &  45 &  37 &  48 &  40 &  39 &  23 &  27 &  26 & 44 &  42 \\
Si I    &  7405.79 &  72 & 107 &  93 & 120 &  92 &  64 &  79 &  83 & 100 & 115 &  86 &  89 &  68 &  76 &  68 & 57 & 100 \\
Si I    &  7415.96 &  69 & 116 &  80 & 132 &  92 &  80 &  78 &  79 &  95 &  94 &  86 &  81 &  63 &  50 &   - & 45 & 100 \\
Ti I    &  6303.77 &   8 &  18 &  22 &  31 &  18 &   - &  18 &  13 &  18 &  39 &  11 &  70 &   5 &  54 &   7 & 13 &  20 \\
Ti I    &  6312.24 &   7 &  17 &  19 &  31 &  17 &   7 &  15 &   - &  20 &  40 &  15 &  61 &   4 &  53 &   3 & 21 &   6 \\
Ti I    &  6336.11 &   6 &  11 &  19 &  25 &  11 &  92 &  93 &   8 &  12 &  35 &   9 &  51 &   5 &  48 &   - &  8 & 105 \\
Ti I    &  6554.22 &  13 &  28 &  34 &   - &  27 &  17 &   - &  19 &  32 &  59 &  17 &  80 &   - &  65 &  14 &  - &   - \\
Ti I    &  6599.13 &   8 &  19 &  26 &  36 &  16 &  10 &  15 &  12 &  23 &  46 &  12 &  65 &   - &  54 &  10 & 26 &   5 \\
Ti I    &  6743.13 &  13 &  36 &  47 &  56 &  33 &  18 &  27 &  19 &  31 &   - &  20 &  88 &  15 &  79 &  15 & 28 & 113 \\
Mg I    &  6319.24 &  21 &  44 &  45 &  65 &  27 &  29 &  25 &  31 &  37 &  76 &  36 &  67 &  17 &  41 &  19 & 26 &  32 \\
Mg I    &  7387.70 &  55 &  87 & 102 & 127 &  76 &  73 &  65 &  74 &  86 & 142 &  82 & 108 &  60 & 105 &  56 & 75 &  35 \\
O I     &  6300.31 &   8 &   5 &   7 &  16 &   4 & 7   &  5  &  9  &   6 &  9  &  10 &  -  &   - &   6 &   - & 11 &  5  \\
\enddata
\end{deluxetable}

\begin{deluxetable}{crrrrrrrrrrrrrrrrrrr}
\label{tb5}
\tabletypesize{\scriptsize}
\tablecolumns{14}
\tablecaption{Measured equivalent widths for the atomic lines: b18 to b35}
\tablewidth{0pt}
\tablehead{
\colhead{Element} & \colhead{$\lambda$} & \colhead{b18} & \colhead{b19} & \colhead{b20} & \colhead{b21} 
& \colhead{b22} & \colhead{b23} & \colhead{b24} & \colhead{b25} & \colhead{b26} &
\colhead{b27} & \colhead{b28} & \colhead{b29} & \colhead{b30} &  \colhead{b31} & \colhead{b32}
& \colhead{b33} & \colhead{b34} & \colhead{b35}
}
\startdata
Ba II  &  4554.03 & 162 &  185 & 138 & 135 & 159 & 162 & 159 & 179 & 150 & 184 & 182 & 167 & 170 & 264 & 178 & 141 & 159 & 142 \\
Ba II  &  4934.10 & 159 &  182 & 147 & 164 & 155 & 153 & 154 & 166 & 152 & 164 & 170 & 161 & 165 & 240 & 168 & 152 & 161 & 147 \\
Ba II  &  6496.93 &  98 &   97 &  82 & 103 &  97 & 106 &  87 & 104 &  85 & 105 & 116 & 102 &  88 & 142 & 111 &  83 &  85 &  80 \\
Zr II  &  6114.85 &   - &    5 &   - &   - &   - &   - &   - &   - &   - &   - &   - &   - &   - &   4 &   - &   - &   - &   - \\
Zr I   &  6134.57 &   - &    - &   - &   - &   - &   - &   - &   6 &   - &   6 &   5 &   - &   8 &   6 &   - &   - &   3 &   - \\
Zr I   &  6140.46 &   - &    - &   - &   - &   - &   - &   - &   2 &   - &   - &   - &   - &   - &   - &   - &   - &   - &   - \\
Zr I   &  6762.40 &   - &    - &   - &   - &   - &   - &   - &   - &   4 &   - &   - &   - &   - &   - &   - &   - &   - &   - \\
La II  &  6390.48 &   7 &    - &   - &   - &   4 &   4 &   3 &   6 &   3 &   6 &   - &   4 &   7 &  10 &   4 &   - &   - &   - \\
La II  &  6774.26 &   8 &    - &   - &   - &   - &   - &   - &   6 &   - &   5 &   7 &   - &   - &   9 &   - &   - &   - &   - \\
Y  II  &  6795.41 &   - &    - &   - &   - &   - &   - &   - &   - &   - &   - &   - &   - &   - &   9 &  13 &   - &   - &   - \\
Eu II  &  6645.10 &   9 &    6 &   7 &   8 &  12 &   7 &   7 &  14 &   7 &   4 &  12 &   7 &   6 &   7 &   9 &  10 &   8 &   6 \\
Ca I   &  6161.30 &  63 &   83 &  75 &  54 &  66 &  66 &  62 &  86 &  69 &  92 &  67 &  57 &  69 &  54 &  50 &  47 &  59 &  61 \\
Ca I   &  6455.60 &  57 &   69 &  48 &  46 &  59 &  55 &  53 &  73 &  54 &  55 &  78 &  42 &  64 &  46 &  46 &  41 &  45 &  46 \\
Ca I   &  6508.85 &  10 &   21 &  10 &   7 &   - &   9 &   8 &  11 &   9 &  22 &  12 &   - &  16 &   6 &  10 &   9 &  10 &   7 \\
Ca I   &  6572.79 &  36 &   53 &  27 &  16 &  33 &  40 &  31 &  76 &  42 &  60 &  28 &   - &  63 &  28 &  20 &  20 &  30 &  30 \\
Si I   &  6237.33 &  55 &   56 &  33 &  51 &  64 &  62 &  57 &  45 &  54 &  70 &  69 &  61 &  43 &  43 &  53 &  48 &   - &  50 \\
Si I   &  6721.84 &  23 &   45 &  28 &  36 &  45 &  47 &  39 &  39 &  36 &  56 &  53 &  40 &  28 &  26 &  43 &  35 &  33 &  31 \\
Si I   &  7226.21 &  35 &   37 &  29 &  30 &  19 &  52 &  36 &  36 &  31 &  45 &   - &  38 &   - &  26 &  41 &  35 &   - &   - \\
Si I   &  7405.79 &  94 &   85 &  67 &  87 &  89 &  91 &  81 &  79 &  85 &  97 &  91 &  90 &  70 &  72 &  94 &  75 &  63 &  82 \\
Si I   &  7415.96 &  78 &   78 &   - &  80 &  84 &  97 &  80 &  79 &  79 &  95 &  89 &  91 &  67 &  71 &  86 &  74 &  60 &  73 \\
Ti I   &  6303.77 &  11 &   18 &  11 &   8 &  10 &  12 &   7 &  31 &  13 &  23 &   7 &   4 &  23 &   6 &   6 &   7 &   7 &  66 \\
Ti I   &  6312.24 &  12 &   19 &  10 &   - & 111 &   6 &   5 &  28 &  12 &  19 &   6 &   4 &  26 &   6 &   - &   5 &   8 &   8 \\
Ti I   &  6336.11 &   8 &   13 &  88 &   - &   8 &   5 &   5 &  20 &   6 &  14 &   5 &   2 &  15 &  85 &   4 &   3 &   7 &   6 \\
Ti I   &  6556.06 &   - &   31 &  17 &  33 &  19 &  18 &  20 &  56 &   - &  41 &   - &   - &   - &   - &   - &   - &  16 &   - \\
Ti I   &  6599.13 &  13 &   20 &  13 &   5 &  12 &  10 &  10 &  38 &  15 &  26 &  27 &   6 &  29 &   9 &   - &   4 &  13 &   8 \\
Ti I   &  6743.13 &  23 &   36 &  16 &   - &  22 &  16 &  20 &  58 &  28 &  47 &   - &  12 &  44 &  21 &   - &   - &  18 &  17 \\
Mg I   &  6319.24 &  34 &   54 &  22 &  20 &  36 &  35 &  31 &  41 &  35 &  46 &  33 &  26 &  36 &  27 &  26 &  23 &  34 &  25 \\
Mg I   &  7387.70 &  71 &   80 &  46 &  62 &  77 &  83 &  73 &  87 &  68 &  99 &  60 &  73 &  66 &   - &   - &   - &  48 &  44 \\
O I    &  6300.31 &   - &   -  &  -  &  4  &  4  &   7 &   3 &  15 &  13 &   6 &  -  &   - &   7 &   5 &   4 &   6 &   - &  6  \\
\enddata
\end{deluxetable}

\begin{deluxetable}{ccccccc}
\label{tb6}
\tablecolumns{5}
\tablewidth{0pc}
\tablecaption{Abundance errors due to stellar parameters uncertainties.}
\tablewidth{0pt}
\tablehead{
\colhead{Star} & \colhead{Element} & \colhead{$\Delta$ T$_{eff}$ = +50 K}
& \colhead{$\Delta$ log$g$ = +0.15 dex} & \colhead{$\Delta\xi$ = +0.10 dex} 
& \colhead{$\Delta$ [Fe/H] = 0.10} & \colhead{$\delta_{t}$}
}
\startdata
b1  &  [Ca/Fe]  &  0.03  &  0.02  &  0.01  &  0.10  &  0.11 \\
  &  [Si/Fe]  &  0.00  &  0.00  &  0.02  &  0.08  &  0.08 \\
  &  [Ti/Fe]  &  0.00  &  0.01  &  0.04  &  0.10  &  0.10 \\
  &  [Mg/Fe]  &  0.02  &  0.03  &  0.00  &  0.09  &  0.10 \\
  &  [Ba/Fe]  &  0.04  &  0.02  &  0.04  &  0.00  &  0.05 \\
  &  [Zr/Fe]  &  0.02  &  0.03  &  0.02  &  0.06  &  0.07 \\
  &  [La/Fe]  &  0.02  &  0.00  &  0.05  &  0.10  &  0.11 \\
  &  [Y/Fe]   &  0.00  &  0.00  &  0.00  &  0.12  &  0.12 \\
  &  [Eu/Fe]  &  0.02  &  0.05  &  0.03  &  0.07  &  0.09 \\
  &  [O/Fe]   &  0.01  &  0.01  &  0.01  &  0.09  &  0.09 \\
b28  &  [Ca/Fe]  &  0.06  &  0.00  &  0.02  &  0.06  &  0.09 \\
  &  [Si/Fe]  &  0.03  &  0.06  &  0.01  &  0.12  &  0.14 \\
  &  [Ti/Fe]  &  0.02  &  0.01  &  0.02  &  0.10  &  0.10 \\
  &  [Mg/Fe]  &  0.02  &  0.03  &  0.02  &  0.14  &  0.14 \\
  &  [Ba/Fe]  &  0.00  &  0.02  &  0.01  &  0.04  &  0.05 \\
  &  [Zr/Fe]  &  0.00  &  0.04  &  0.00  &  0.03  &  0.05 \\
  &  [La/Fe]  &  0.05  &  0.04  &  0.00  &  0.04  &  0.07 \\
  &  [Y/Fe]   &  0.00  &  0.05  &  0.05  &  0.09  &  0.11 \\
  &  [Eu/Fe]  &  0.03  &  0.15  &  0.13  &  0.10  &  0.22 \\
  &  [O/Fe]   &  0.00  &  0.03  &  0.00  &  0.06  &  0.07 \\
\enddata
\end{deluxetable}


\clearpage

\clearpage

\clearpage

\clearpage

\clearpage

\clearpage

\clearpage

\end{document}